# Soft Coulomb Gap Limits the Performance of Organic Thermoelectrics

*Yuqian Liu[1], Xiaoran Wei[1], Dorothea Scheunemann[2], Maojie Zhang[3], Wanlu Zhang[1], Martijn Kemerink[2], and Guangzheng Zuo[1]\**

[1]Institute for Electric Light Sources, College of Intelligent Robotics and Advanced Manufacturing, Fudan University, 200433, Shanghai, P. R. China

[2]Institute for Molecular Systems Engineering and Advanced Materials, Heidelberg University, 69120, Heidelberg, Germany

[3]National Engineering Research Center for Colloidal Materials, Key Laboratory of Special Functional Aggregated Materials, Ministry of Education, School of Chemistry & Chemical Engineering, Shandong University, 250100, Jinan, Shandong, P. R. China

\* Corresponding author, email: gzzuo@fudan.edu.cn

# Abstract:

Although consensus exists that the thermoelectric properties of doped organic semiconductors result from a complex interplay between a large number of mutually dependent factors, there is no consensus on which of these are dominant, or even on how to best describe the charge and energy transport. This holds particularly in the intermediate doping regime where the optimal performance is typically observed at the roll-off in the Seebeck coefficient - conductivity $(S - \sigma)$ correlation, fundamentally limiting the rational advancement of organic thermoelectric materials. Here, we combine experiments on a board set of conjugated polymers with kinetic Monte Carlo simulations across varying doping levels to uncover a general transport framework. We demonstrate that the optimal thermoelectric power factor $(PF_{max})$ consistently occurs at the transition between conventional variable-range hopping (VRH) and VRH in a density of states in which a soft Coulomb gap forms at the Fermi level, as described by Efros and Shklovskii (ES-VRH). This suggests the use of high dielectric constant materials or the promotion of charge delocalization as an avenue to shift the roll-off of the $S - \sigma$ curve, which constrains $PF_{max}$, to higher doping levels and accordingly higher $PF$.

# Introduction

Conjugated polymers (CPs) have garnered increasing attention as promising organic thermoelectric (OTE) materials due to their solution processability, mechanical flexibility and intrinsically low thermal conductivity.[1–6] Commonly, the performance of thermoelectric materials is evaluated using the dimensionless Fig. of merit $ZT$, given by $ZT = \frac{\sigma S^2}{\kappa}T$, where $\sigma$ is the electrical conductivity, $S$ is the Seebeck coefficient, $\kappa$ is the thermal conductivity and $T$ is the temperature. Over the last decade, research on organic OTEs has advanced $ZT$ to values around unity at room temperature[7], a level nearly equivalent to that of their inorganic counterparts. However, as Brunetti et al. pointed out [8], the field of OTEs appears to be stuck in terms of power factor ($PF = \sigma S^2$), and new approaches for rational material design are urgently needed. This bottleneck can be attributed to a limited comprehensive understanding of the complex interplay of numerous interrelated factors. Among these, the relationship between the doping level, charge transport mechanisms, and the resulting thermoelectric performance is particularly critical, yet poorly understood.

Owing to the disordered nature and ultra-low charge concentration of intrinsic CPs, a common and effective strategy to enhance their electrical conductivity is molecular doping, which elevates the carrier density but often suppresses $S$.[9–12] Hence, striking a balance between $\sigma$ and $S$ is essential for achieving high-efficiency OTEs. When doped over a sufficient range, a transition regime typically occurs in the $S$–$\sigma$ correlation[13,14]. At low doping concentrations until the maximum power factor, $PF_{max}$, the entire thermopower versus conductivity curve can be described by the well-known empirical power law $S \propto \sigma^{-0.25}$ as identified by Glaudell et al[15]. The latter can be attributed to transport by VRH in a density of states (DOS) that is modified by Coulomb traps formed by ionized dopants[16]. In the intermediate doping regime, Li et al. indicated that the DOS is renormalized due to the overlapping Coulomb potentials of homogeneously distributed ionized dopants.[17] In consequence, transport at these dopant concentrations can, in stark contrast to low-doping concentrations, be described as transport of free charge carriers, not being limited by charge trapping by the Coulomb potentials of the dopant counterions. Importantly, the maximum power factor is typically achieved in this regime[14,18]. Beyond $PF_{max}$, the $S$-$\sigma$ correlation instead follows $S \propto ln(\sigma)$, as also observed in experiments[12,15,19]. Recently, Yang et al. indicated the Seebeck coefficient follows Heike's formula at intermediate to high doping levels and attributed this to the conductivity becoming limited by carrier–carrier repulsion.[12] However, the underlying physics remained unclear, and a comprehensive understanding of how the transition affects the optimal $PF$ is lacking.

Here, we combine experimental measurements and kinetic Monte Carlo simulations to elucidate the fundamental mechanisms governing the Seebeck coefficient–conductivity $(S - \sigma)$ correlation in doped conjugated polymers across the entire doping range. We demonstrate that the transition from the empirical power law $S \propto \sigma^{-0.25}$ to the logarithmic dependence $S \propto ln(\sigma)$ universally originates from the formation of a soft Coulomb gap in the renormalized Gaussian density of states, reflecting the dominance of carrier–carrier Coulomb interactions.[12,20] Although this transition is universally observed, its precise manifestation depends critically on the material's dielectric properties and energetic disorder. Shifting the $S - \sigma$ roll-off to higher doping levels is identified as a key strategy for enhancing the maximum power factor. Accordingly, we propose that enhancing dielectric screening by increasing the relative permittivity and promoting charge delocalization to mitigate Coulomb interactions represent effective design strategies for high-performance organic thermoelectric materials. These findings establish a robust mechanistic framework for rationally tailoring electronic structure and charge transport to optimize thermoelectric performance.

## Results and Discussion

The materials investigated in this study are depicted in Fig. 1. To enable a comprehensive and systematic exploration of the structure–property relationship governing thermoelectric performance, we selected a series of conjugated polymers (CPs) featuring diverse conjugated backbones and covering a wide range of HOMO energy levels from -5.0 eV to -5.5 eV. These CPs include distinct molecular design strategies, including donor–donor (D–D), donor–acceptor (D–A), acceptor–acceptor (A–A) motifs, as well as block copolymer architectures, thereby offering a broad platform for correlating molecular structure with charge transport and thermoelectric behavior. The thermoelectric properties for all materials were evaluated by measuring the conductivity and Seebeck coefficient on lateral (in-plane) devices, following procedures outlined in detail in our previous works[18,21,22] or adapted from relevant literature sources[23–35]. $FeCl_3$, with a LUMO below -5.8 eV[36,37], was employed as the (p-type) dopant for all polymers in this study. More details are provided in the Methods Section. All the thermoelectric properties are shown in Fig. S2-S5 and summarized Tabel S2 in the Supporting Information.

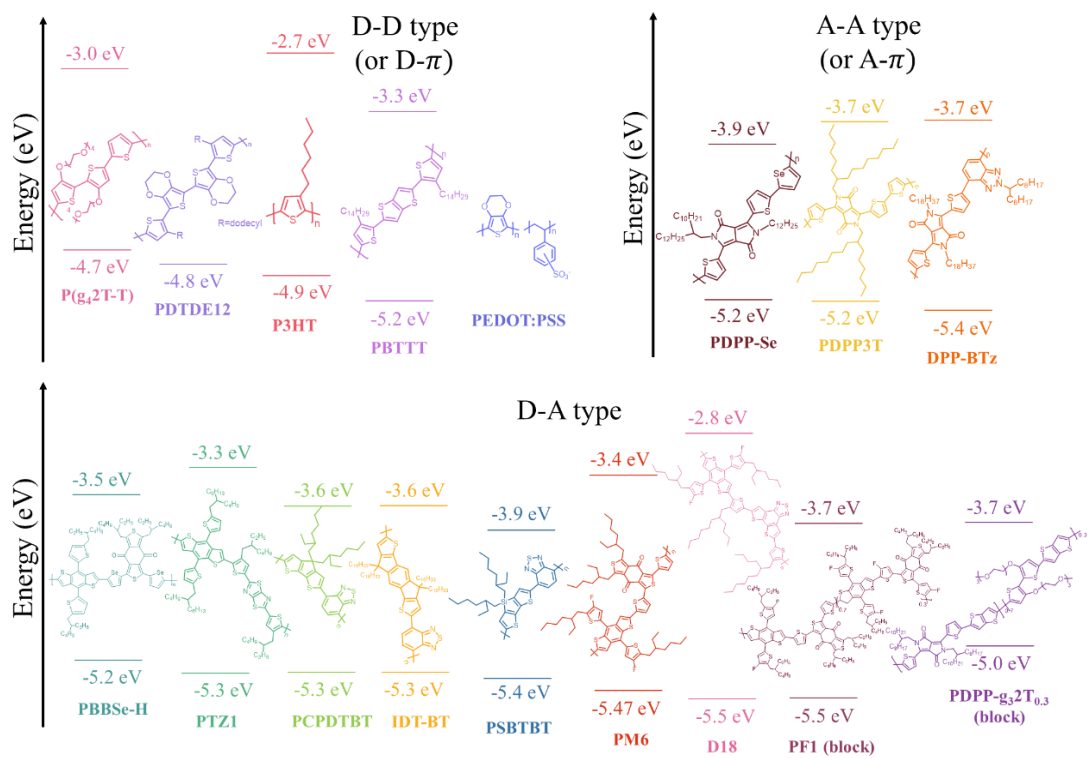

**Fig. 1 | Chemical structure and energy levels of the studied materials.** Molecular structures and the corresponding highest occupied molecular orbital (HOMO) and lowest unoccupied molecular orbital (LUMO) energy levels of the conjugated polymers investigated in this study

To better understand the relationship between chemical structure and thermoelectric performance and its influence on charge transport, we introduce a new representation of thermoelectric parameters, in which both $S$ and $\sigma$ are scaled relative to their respective values at maximum power factor, denoted as $S_{PFmax}$ and $\sigma_{PFmax}$, respectively. The resulting trends are shown in Fig. 2. It reveals that the Seebeck coefficient–conductivity ($S-\sigma$) curves universally follow the well-known empirical relationship $S \propto \sigma^{-0.25}$ in the low conductivity, i.e. low-doping, regime. This behavior was initially identified by Glaudell et al,[15] and the underlying physical mechanism was subsequently elucidated by Abdalla et al.,[16] who attributed it to charge transport being predominantly governed by an exponential tail of deep states in the intrinsically Gaussian DOS due to the attractive Coulomb potential surrounding ionized dopants. However, beyond a certain point, this relationship transitions to a logarithmic dependence, $S \propto ln(\sigma)$, indicating a shift in the dominant charge transport mechanism. In addition, the $S \propto ln(\sigma)$ trend exhibits a strong material dependence, and as such is not universal.

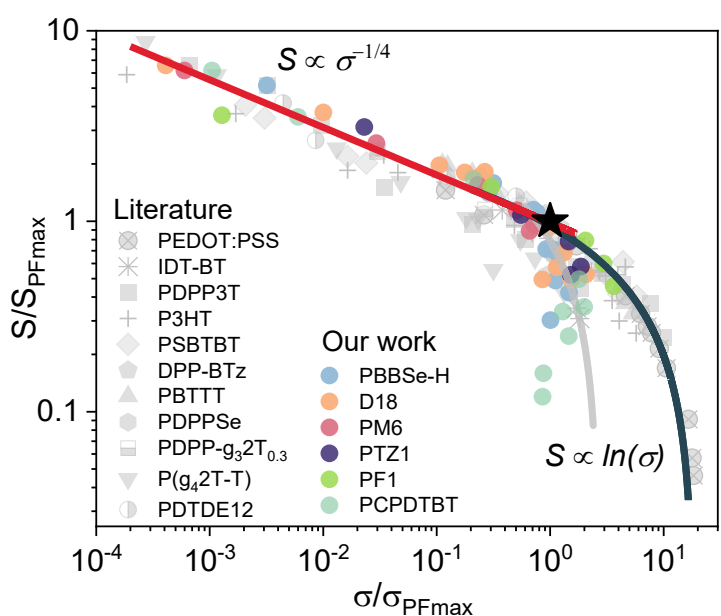

**Fig. 2 | Normalized Seebeck coefficient versus conductivity.** Both the Seebeck coefficient ($S$) and conductivity ($\sigma$) are normalized to their respective values at maximum power factor. The star symbol represents the maximum PF ($PF_{max}$). ⊗ PEDOT:PSS [26], ✶ IDT-BT [27], ■ PDPP3T [28], + P3HT [29], ◆ PSBTBT [29], ⬠ DPP-BTz [30], ▲ PBTTT [31], ⬢ PDPPSe [32], ▭ PDPP-$g_3$2T$_{0.3}$ [33], ▼ P($g_4$2T-T) [34], ◐ PDTDE12[35].

To elucidate the origin of the transition regime in the $S-\sigma$ correlation, we conducted temperature-dependent conductivity measurement on lateral (in-plane) devices for all studied materials across various doping concentrations and temperature gradients. The temperature dependence of the electrical conductivity in a disordered system can generally be described by:

$$\sigma = \sigma_0 \exp\left[-\left(\frac{T_0}{T}\right)^\alpha\right] \quad (1)$$

where $\sigma_0$ denotes the prefactor, and $T_0$ is the characteristic temperature. Both $\sigma_0$ and $T_0$ are material- and film-dependent parameters influenced by factors such as the electronic density of states, Debye frequency of the material, localization length and dielectric constant.[38] The exponent $\alpha$ determines the (stretching of the) temperature dependence of the conduction and depends on the charge transport mechanism. Specifically, for a more or less constant DOS around the Fermi level, $\alpha = 1/(1+d)$ with $d$ the dimensionality of the system, for 3D systems leading to Mott-type variable range hopping (Mott-VRH) with $\ln \sigma \propto T^{-0.25}$. For systems that are not evidently 2D, the exponent $\alpha = 0.5$ is typically attributed to the opening of a soft Coulomb gap at the Fermi energy, resulting in Efros-Shklovskii variable-range hopping (ES-VRH).[39]

As shown in Fig. 3a, the conductivity of the material PTZ1 does not exhibit a good linear fit in the low doping regime. In contrast, the conductivity versus temperature curve shows an excellent linear trend, that is, $\alpha = 0.5$, starting at a doping concentration of ~15 mM $FeCl_3$, which corresponds to the maximum PF ($PF_{max}$) value. This indicates that the charge transport follows the ES-VRH model at and beyond the $PF_{max}$. Interestingly, prior to $PF_{max}$, the conductivity versus temperature curve exhibits an excellent linear trend with an exponent of 0.25, as shown in Fig. S6 of the Supporting Information, suggesting conventional Mott-VRH behavior. In that representation, deviations from a linear trend are observed at higher doping concentrations. Hence, for the material PTZ1, across the doping regime from 5 mM to 50 mM and the temperature range from 90 K to 300 K, a clear change in the exponent for the relation $\ln(\sigma) \propto T^{-n}$ is observed, shifting from 0.25 to 0.5 across the entire doping concentration range, with the transition occurring at the doping concentration of $PF_{max}$, as shown in the inset and, in more detail, in Fig. S12. The change in exponent indicates a transition in the charge transport mechanism from Mott-VRH to ES-VRH. A similar trend was observed for the material PBBSe-H, as shown in Fig. 3b and Fig. S7, S13, i.e. a linear trend with an exponent of 0.25 at low doping concentration, which shifts to 0.5 starting at the doping concentration of 7 mM FeCl$_3$, corresponding to $PF_{max}$. Similar observations of the exponent transitioning from 0.25 to 0.5 at $PF_{max}$ were found in the other materials systems, as depicted in Fig. S8 to S11 and S14 to S17. This suggests a rather general transition in the charge transport mechanism from Mott-VRH to ES-VRH for doped CPs, with the transition consistently occurring around $PF_{max}$. Simultaneously, the $S - \sigma$ correlation changes from the well-known empirical power law $S \propto \sigma^{-0.25}$ to $S \propto \ln(\sigma)$, as shown in Fig. 2.

To further investigate the universality of the observed transition in the charge transport mechanism from Mott-VRH to ES-VRH for doped CPs, kinetic Monte Carlo (kMC) simulations were performed. The employed model has been described in detail in previous studies.[21,40–42]. In brief, it gives a numerically exact description of Coulombically interacting localized particles, hopping on a fixed lattice with site energies drawn from a Gaussian DOS. The input parameters used in the simulations were, where possible, derived from experimental data of the materials PTZ1 and PBBSe-H. The calculated temperature-dependent conductivity as a function of charge carrier concentration is presented in Fig. 3c and 3d. These results are consistent with the experimental data, showing that the conductivity versus temperature curve exhibits an excellent linear trend with an exponent of 0.5 at and beyond the concentration of optimal $PF$ (~10%). This finding further confirms the universal charge transport trend in doped CPs.

The kMC simulations also reproduce the lower exponents (<0.5) at lower doping levels, as shown in Fig. 3c (insert) and Fig. S19 in the SI. However, the upswing in exponent at the lowest doping levels is not visible in the present experiments. We attribute this to the fact that all simulations are performed for a strictly Gaussian DOS, as is appropriate in the transition regime and beyond, as this is our region of interest.[17] At lower concentrations, the DOS shape is more complicated due to the dopant-induced trap states, as discussed in the introduction.[16] Moreover, the experiments may simply not reach these low doping levels.

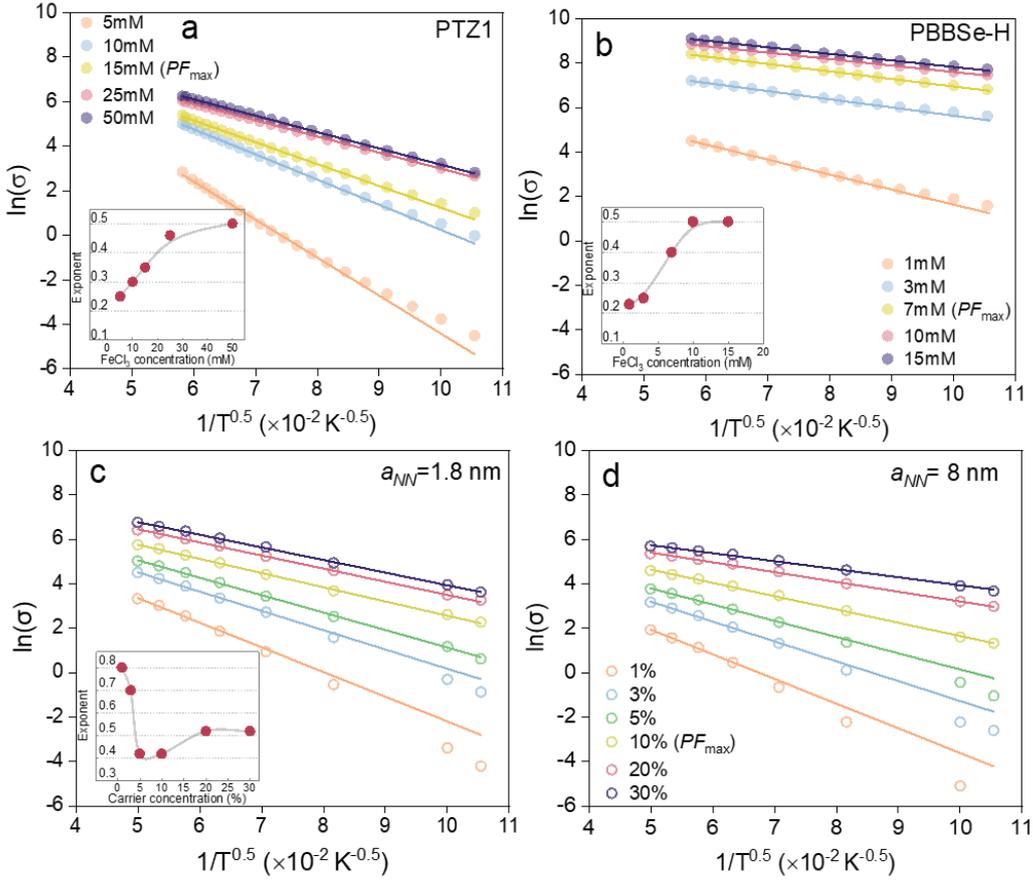

**Fig. 3 | Temperature-dependent conductivity in experiments and simulations.** Logarithmic conductivity as a function of temperature, varying with $FeCl_3$ (dopant) concentration for experiments: **a** PTZ1, **b** PBBSe-H, and kinetic Monte calculations (kMC): **c** with inter-site distance $a_{NN}$= 1.8 nm and **d** $a_{NN}$= 8 nm. Default kMC parameters: energetic disorder $\sigma_{DOS}$= 75 meV, attempt-to-hop frequency $\nu_0$= 1×10$^{13}$ s$^{-1}$, relative dielectric constant $\varepsilon_r$= 3.6, temperature $T$ from 100 K to 400 K. Lines are fits to $\ln \sigma \propto T^{-0.5}$.

Before discussing the possible underlying reasons for the transition in the charge transport mechanism from Mott-VRH to ES-VRH in doped CPs with increasing doping concentration, we will first clarify the differences between Mott-VRH and ES-VRH in disordered organic systems. Both models describe charge transport in strongly disordered systems via hopping between localized states. However, they differ in the physical mechanisms that are accounted for and hence have different conditions for applicability, primarily depending on whether Coulomb interactions play a significant role and how these interactions affect the DOS. Specifically, ES-VRH emerges when strong Coulomb repulsion, arising from carrier-carrier interactions, suppresses the DOS near the Fermi level, leading to the formation of a soft Coulomb gap, which subsequently governs the charge transport behavior. Moreover, temperature has a significant influence on the soft Coulomb gap. As temperature increases, charge carriers gain sufficient thermal energy to overcome the Coulombic suppression near the Fermi level, resulting in a washing-out of the gap and a corresponding restoration of the original DOS. Consequently, the width and depth of the gap decrease with increasing temperature, thereby diminishing its influence on charge transport.

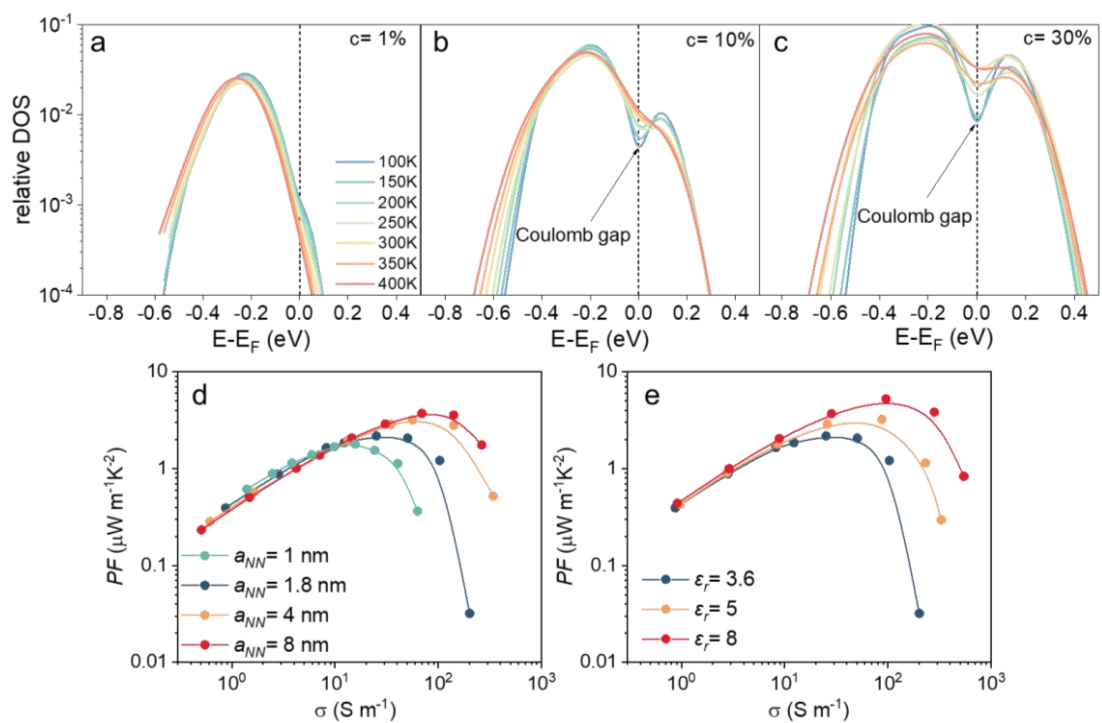

**Fig. 4 | Kinetic Monte Carlo simulations of DOS and thermoelectric behavior.** Kinetic Monte Carlo calculation (kMC) results for the evolution of density of states (DOS) as a function of temperature at different hole concentrations for **a** 1%, **b** 10% and **c** 30%. The change of thermoelectric characteristics by kMC simulation with different parameters for **d** varying inter-site distance $a_{NN}$= 1 nm, 1.8 nm, 4 nm and 8 nm, and **e** varying relative dielectric constant $\varepsilon_r$= 3.6, 5 and 8. The used (default) kMC

parameters are: $a_{NN}$ =1.8 nm, $\sigma_{DOS}$ = 75 meV, $\nu_0$ = 1×10$^{13}$ s$^{-1}$, $\varepsilon_r$ = 3.6. For DOS calculation in panel a-c, the same input parameters as in Fig. 3c were used.

To verify whether indeed the observed change in exponent is due to a transition in transport from Mott-VRH to ES-VRH, in turn originating from the formation of soft Coulomb gap with increasing doping concentration, the detailed evolution of the DOS with temperature was investigated by kMC simulations. As shown in Fig. 4, for a carrier concentration $c$ = 1%, there is negligible change in DOS near the Fermi energy as the temperature increases from 100 K to 400 K. However, when $c$ increases to 10%, a distinct parabolic dip appears in the DOS near $E_F$ around ~200 K, indicating the formation of a soft Coulomb gap. As the temperature increases to 400 K, this soft Coulomb gap gradually disappears. Combining these observations with the temperature-dependent conductivity behavior shown in Fig. 3c, it is evident that at $c$ = 1%, no soft Coulomb gap is formed, and the charge transport is dominated by the Mott-VRH model, characterized by $ln(\sigma) \propto T^{-0.25}$. As $c$ increases to 10%, the carrier-carrier repulsions become significant, leading to the formation of soft Coulomb gap and a corresponding transport behavior of $ln(\sigma) \propto T^{-0.5}$. With further increase in $c$ up to 30%, the soft Coulomb gap becomes even more pronounced, as shown in Fig. 4c.

The observed transition from Mott-VRH to ES-VRH, indicated by the exponent shift from –0.25 to –0.5, appears to be a generic phenomenon across various doped conjugated polymers. Our kMC simulations suggest this universality arises because of the formation of a soft Coulomb gap and thus the onset of ES-VRH primarily depend on a small set of fundamental parameters: specifically, the density of states (DOS) near the Fermi level, determined indirectly by the nearest-neighbor distance ($a_{NN}$), the width of the Gaussian DOS ($\sigma_{DOS}$), and the material's dielectric constant ($\varepsilon_r$). Importantly, these parameters typically exhibit only algebraic variations across different polymer systems, rather than exponential changes. Consequently, a similar DOS renormalization profile—and thus a soft Coulomb gap—is consistently expected to emerge within a doping range of ~ 5-20%. To further substantiate this point, we performed additional simulations exploring DOS evolution for different $a_{NN}$, $\sigma_{DOS}$ and $\varepsilon_r$ values. The results (shown in Fig. S20 to S23 of the Supporting Information) confirm that the soft Coulomb gap consistently appears within this doping regime, implying that the ES-VRH transport regime should generally be observable provided sufficient doping levels can be experimentally realized. Indeed, our experimental data aligns well with this prediction, clearly exhibiting a transition from exponent –0.25 to –0.5 whenever doping extends sufficiently beyond $PF_{max}$.

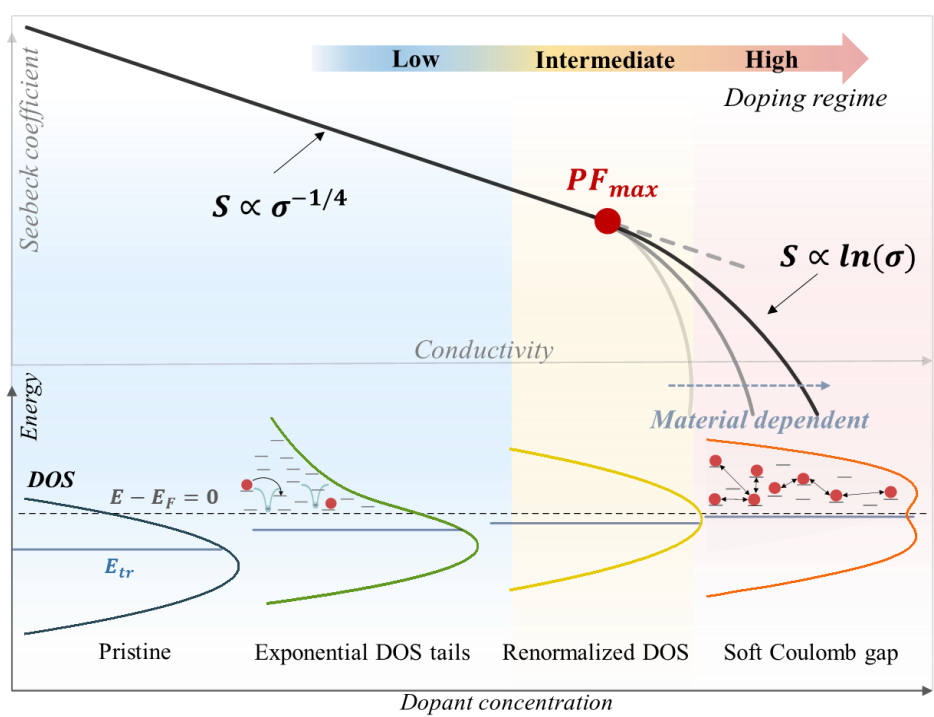

**Fig. 5 | Doping-driven evolution of the DOS and the Seebeck–conductivity trend.** The evolution of the shape of the density of states (DOS) progresses from the intrinsic Gaussian distribution to the formation of exponential tails, then transitions into a renormalized DOS, and ultimately develops into a Coulomb-gap structure with increasing doping concentration, and the corresponding $S - \sigma$ correlation.

Finally, to put our results in perspective, we illustrate the interrelationship between the density of states, charge transport and thermoelectric properties across the entire doping regime. In general, the detailed characteristics of the DOS determine charge transport in disordered organic systems, and thereby play a dominant role in shaping the corresponding thermoelectric properties. Fig. 5 illustrates the evolution of the DOS profile and the associated $S - \sigma$ correlation with increasing doping concentration. As discussed in the introduction, at low doping concentrations, the overall DOS, featuring an additional exponential tail, introduced by dopant ions, below an otherwise Gaussian DOS governs charge transport. In this regime, charge carriers follow the conventional VRH mechanism, resulting in the empirical power law of $S \propto \sigma^{-0.25}$. With increasing doping concentration, up to intermediate levels, the overlapping Coulomb potentials of uniformly distributed ionized dopants progressively renormalize the DOS, enabling (effectively) free-carrier transport rather than trap-limited hopping. Notably, the optimal power factor is typically achieved in this regime. Unlike in the previous regime, the Fermi and transport energies lie in a region where the DOS is no longer strongly energy dependent, leading to Mott-like VRH (Eq. 1 with $\alpha \approx 1/4$). As the doping concentration continues to increase, the carrier-carrier interactions become significantly more pronounced, leading to the formation of a soft Coulomb gap around the Fermi level in the renormalized DOS, which governs the charge transport and induces a transition to the ES-VRH model (Eq. 1 with $\alpha \approx 1/2$). Consequently, the $S - \sigma$ correlation shifts to a logarithmic dependence, described by $S \propto ln(\sigma)$. It is worth noting that the $S \propto ln(\sigma)$ relationship is strongly material-dependent, i.e., it is not universal and is significantly influenced by the degree of disorder in the material.

The discussion can be translated into a design rule for reaching high performance in OTE materials. Since $PF \propto \sigma^{1/2}$ till the roll-off point at max PF, one should aim to shift the roll-off to higher doping levels. This can be accomplished by suppressing carrier–carrier interactions to reduce the impact of the soft Coulomb gap on thermoelectric performance. Since the Coulomb interaction scales as $E_C = e/4\pi\varepsilon_r\varepsilon_0 a$), two strategies to achieve this can be identified, as depicted in Fig. 4d and e (and Fig. S24 in SI). First, by increasing the relative permittivity ($\varepsilon_r$) of the material, for example by introducing polarizable side chains, and, second, by promoting charge delocalization, e.g. by planarizing the backbone, to increase the effective separation between carriers.

Moreover, the emergence of a soft Coulomb gap at high doping levels is a general phenomenon in disordered organic semiconductors. Thus, the insights gained here are not limited to organic thermoelectrics but are broadly relevant to other doping-driven applications, including organic light-emitting diodes (OLEDs), organic field-effect transistors (OFETs), and other emerging devices.

## Conclusion

In summary, we have conducted a comprehensive study by combining experimental measurements with kinetic Monte Carlo simulations to elucidate the fundamental mechanisms governing the correlation between the Seebeck coefficient and electrical conductivity in doped conjugated polymers. Our findings reveal that the observed transition from the empirical power law $S \propto \sigma^{-0.25}$ to the logarithmic dependence $S \propto ln(\sigma)$ originates from the formation of a soft Coulomb gap in the renormalized Gaussian density of states (DOS). This transition reflects the dominance of carrier–carrier Coulomb interactions. Furthermore, we demonstrate that although this transition is universally observed across various material systems, its exact shape and the position of the $S - \sigma$ roll-off are critically dependent on the material's dielectric properties and degree of energetic disorder. These findings offer critical mechanistic insights into the interplay between electronic structure, charge transport, and thermoelectric performance, establishing a solid foundation for the rational design of next-generation high-performance organic thermoelectric materials. Specifically, the ability to shift the roll-off to higher doping concentrations is identified as a key factor for achieving enhanced power factors. Based on these insights, we propose a design strategy for developing high-performance OTE materials minimizing Coulomb interactions by either increasing the relative permittivity or by promoting charge delocalization.

## Methods

**Materials:** The polymers D18, PM6, and PCPDTBT were purchased from Solarmer Materials Inc. PTZ1, PBBSe-H, and PF1 were synthesized and provided by Zhang's group[43,44]. The solvent acetonitrile (AR, ⩾99.0%) was purchased from Sinopharm Chemical Reagent Co., Ltd., and chlorobenzene (AR, ⩾99.5%) from Shanghai Titan Scientific Co Ltd. All materials were used as received without further purification.

**Device fabrication:** The glass substrates with patterned indium tin oxide (ITO) electrodes were sequentially cleaned by sonication in soapsuds, deionized water, acetone, ethyl alcohol, and isopropanol, followed by drying under a nitrogen stream. Polymer solutions (10 mg ml$^{-1}$ in chlorobenzene) were spin-coated at 2000 rpm for 45 seconds, and then 3000 rpm for 25 seconds, followed by thermal annealed at 160 °C for 20 minutes. For doped polymer thin films, the pre-annealed films were immersed in FeCl$_3$/acetonitrile solution of varying concentrations for 60 seconds, followed by a spin-drying to remove the residual solution. The films were then rinsed with acetonitrile and annealed at 60 °C for 10 minutes to remove residual solvent. The final film thickness ranged from 40 to 60 nm, as measured by a Dektak surface profilometer.

**Conductivity and Seebeck coefficient measurement:** The devices were fabricated with a lateral (in-plane) structure. All measurements were performed immediately after device fabrication in a glove box under a dry nitrogen atmosphere at room-temperature. Current-voltage characteristics were obtained between −50 and 50 mV. The Seebeck coefficient was determined by applying a controlled temperature gradient across the sample using $S = \Delta V/\Delta T$. Additional experimental details can be found in our previous work.[18,21,22] For temperature-dependence of conductivity measurement, the *I-V* characteristics were recorded over a voltage range of −50 to 50 mV, with temperature varied from 90 K to 300 K in 10 K increments. All measurements were performed in a custom-built, light-blocking vacuum chamber (P < 0.1 mbar). At each temperature step, the device was allowed to thermally equilibrate for 30 minutes prior to measurement. Current readings were then acquired using a Keithley 2450 SourceMeter. For each doping concentration, measurements were conducted on approximately 3-5 devices to ensure reproducibility.

## Data availability

The data supporting the findings of this study are included within the article and its Supplementary Information.

## Conflict of Interest

The authors declare no conflict of interest.

## Acknowledgements

This work was supported by the Science and Technology Commission of Shanghai Municipality (STCSM) under grant No. 23ZR1407400. This work is financially supported by the European Commission through the Marie Sklodowska-Curie project HORATES (GA-955837). M.K. thanks the Carl Zeiss Foundation for financial support. G.Z. thanks Xiaoliang Mo for the use of technical infrastructure for device fabrication.

# Supplementary Information

**Soft Coulomb Gap Limits the Performance of Organic Thermoelectrics**

*Yuqian Liu[1], Xiaoran Wei[1], Dorothea Scheunemann[2], Maojie Zhang[3], Wanlu Zhang[1], Martijn Kemerink[2], and Guangzheng Zuo[1]\**

[1]Institute for Electric Light Sources, College of Intelligent Robotics and Advanced Manufacturing, Fudan University, 200433, Shanghai, P. R. China

[2]Institute for Molecular Systems Engineering and Advanced Materials, Heidelberg University, 69120, Heidelberg, Germany

[3]National Engineering Research Center for Colloidal Materials, Key Laboratory of Special Functional Aggregated Materials, Ministry of Education, School of Chemistry & Chemical Engineering, Shandong University, 250100, Jinan, Shandong, P. R. China

\* Corresponding author, email: gzzuo@fudan.edu.cn

# Content

S1-Full names, chemical structures, and energy levels of all studied polymers

D-D (or D-π) type :

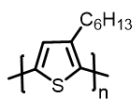
P3HT

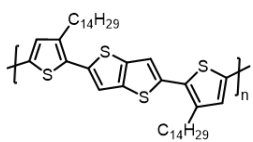
PBTTT

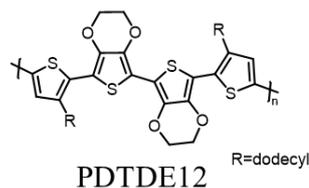
PDTDE12    R=dodecyl

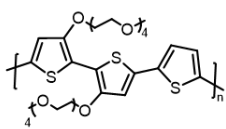
P(g$_4$2T-T)

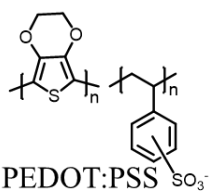
PEDOT:PSS  SO$_3^-$

A-A (or A-π) type:

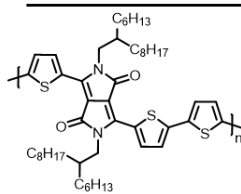
PDPP3T

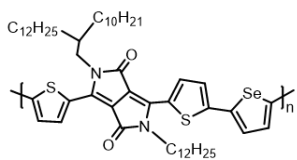
PDPP-Se

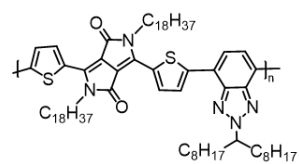
DPP-BTz

D-A type :

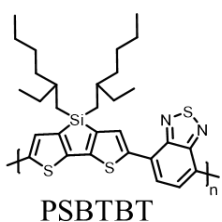
PSBTBT

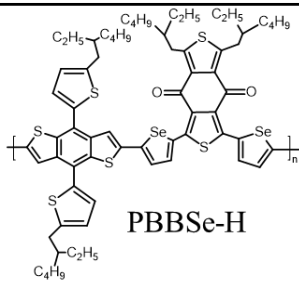
PBBSe-H

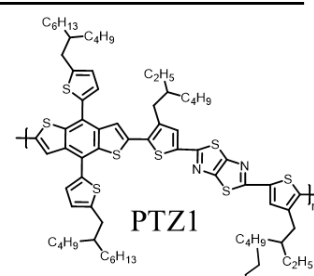
PTZ1

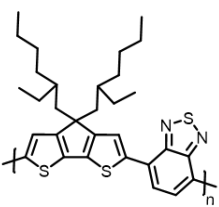
PCPDTBT

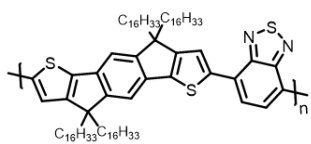
IDT-BT

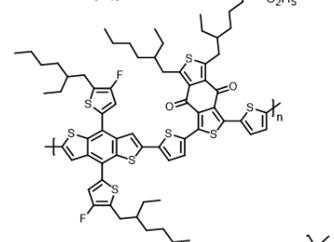
PM6

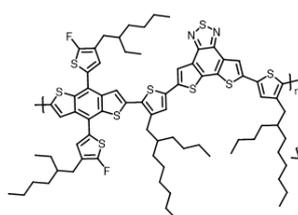
D18

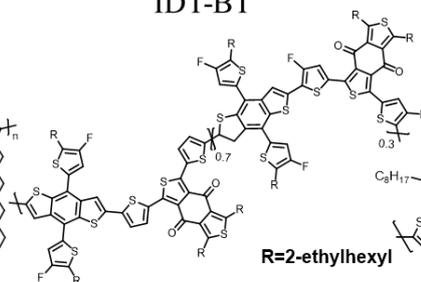
PF1    R=2-ethylhexyl

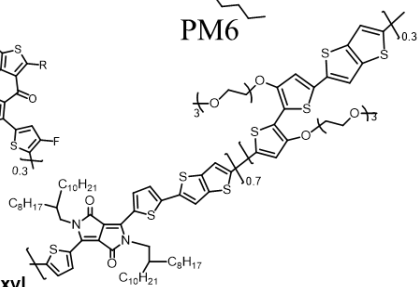
PDPP-g$_3$2T$_{0.3}$

**Figure S1** | Chemical structure of all used polymers. The Full names of the polymers are listed below.

**P3HT:** Poly(3-hexylthiophene).
**PBTTT:** Poly[thieno[3,2-b]thiophene-2,5-diyl(4,4'-ditetradecyl[2,2'-bithiophene]-5,5'-diyl)].
**PDTDE12:** Poly[1,4-phenylene-2,5-bis(5-(3-(dodecylsulfanyl)thiophen-2-yl)-2,3-dihydrothieno[3,4-b][1,4]dioxine-5-yl)].
**P(g$_4$2T-T):** Poly[3,3′-bis(3,6,9,12-tetraoxatridec-1-yloxy)[2,2′:5′,2″-terthiophene]-5,5″-diyl].
**PEDOT: PSS:** Poly(3,4-ethylenedioxythiophene): poly(styrene- sulfonate).
**PDPP3T:** Poly[{2,5-bis(2-hexyldecyl)-2,3,5,6-tetrahydro-3,6-dioxopyrrolo[3,4-c]pyrrole-1,4-diyl}-alt-{[2,2'v:5'2"terthiophene]-5,5"-diyl}].
**PDPP-Se:** Poly(diketopyrrolopyrrole-co-selenophene).
**DPP-BTz:** Poly[[2-(1-octylnonyl)-2H-benzotriazole-4,7-diyl]-2,5-thiophenediyl[2,3,5,6-tetrahydro-2,5-dioctadecyl-3,6-dioxopyrrolo[3,4-c]pyrrole-1,4-diyl]-2,5-thiophenediyl].
**PSBTBT:** Poly[(4,4'-bis(2-ethylhexyl)dithieno[3,2-b:2',3'-d]silole)-2,6-diyl-alt-(2,1,3-benzothiadiazole)-4,7-diyl].
**PBBSe-H:** Poly[4,8-bis(5-(2-ethylhexyl)thien-2-yl)benzo[1,2-b:4,5-b′]dithiophene-alt-5,6-bis(4-(selenophen-2-yl)thiophen-2-yl)-2,1,3-benzothiadiazole-4,7-dione].
**PTZ1:** Poly[thiazolo[5,4-d]thiazole-2,5-diyl[4-(2-ethylhexyl)-2,5-thio-phenediyl][4,8-bis[5-(2-butyloctyl)-2-thienyl]benzo[1,2-b:4,5-b']dithiop-hene-2,6-diyl][3-(2-ethylhexyl)-2,5-thiophenediyl]].
**PCPDTBT:** Poly[2,6-(4,4-bis-(2-ethylhexyl)-4H-cyclopenta[2,1-b;3,4b']-dithiophene)-alt-4,7-(2,1,3-benzothiadiazole)].
**IDT-BT:** Poly[2,1,3-benzothia-diazole-4,7-diyl(4,4,9,9-tetrahexadecyl-4,9-dihydro-s-indaceno[1,2-b:5,6-b']dithiophene-2,7-diyl)].
**PM6:** Poly((4,8-bis(5-(2-ethylhexyl)-4-fluoro-2-thienyl)benzo[1,2-b:4,5-b′]dithiophene-2,6-diyl)-2,5-thiophenediyl-(5,7-bis(2-ethylhexyl)-4,8-dioxo-4H,8H-benzo[1,2-c:4,5-c']dithio-phene-1,3-diyl)-2,5-thio-phenediyl).
**D18:** Poly(dithieno[3,2-e:2',3'-g]-2,1,3-benzothiadiazole-5,8-diyl(4-(2-buty-loctyl)-2,5-thiophenediyl)(4,8-bis(5-(2-ethylhexyl)-4-fluoro-2-thienyl)benzo[1,2-b:4,5-b'] dithiophene-2, 6-diyl) (3-(2-butyloctyl)-2,5-thiophenediyl)).
**PF1:** Poly[(5,6-difluoro-2,1,3-benzothiadiazole-alt-4,8-bis(5-(2-ethylhexyl) thiophen-2-yl)benzo[1,2-b:4,5-b′]dithiophene)-ran-(1,3-bis(5-bromo-4-fluorothiophen-2-yl)-5,7-bis(2-ethylhexyl)-4H,8H-benzo[1,2-c:4,5-c']bisthiophene-4,8-dione)].
**PDPP-g$_3$2T$_{0.3}$:** Poly[2,5-bis(2-octyldodecyl)-3,6-bis(thiophen-2-yl)pyrrolo[3,4-c]pyrrole-1,4(2H,5H)-dione-alt-3,3'-bis{2-[2-(2-methoxyethoxy)ethoxy]ethoxy}-2,2'-bithiophene].

Table S1 Energy levels for the polymers studied in this paper, collected from the literature. We note that the absolute energy level values for each polymer may vary slightly across different sources. However, all polymers are expected to exhibit efficient p-type doping with the commonly used dopant $FeCl_3$, which has a LUMO level below -5.8 eV.

| Materials | HOMO (eV) | LUMO (eV) | Reference |
| --- | --- | --- | --- |
| P3HT | -4.9 | -2.7 | Ref. 1 |
| PSBTBT | -5.37 | -3.88 | Ref. 2 |
| PDPP3T | -5.18 | -3.68 | Ref. 3 |
| IDT-BT | -5.34 | -3.62 | Ref. 4 |
| PBTTT | -5.24 | -3.35 | Ref. 5 |
| DPP-BTz | -3.65 | -5.37 | Ref. 6 |
| PDPP-Se | -5.15 | -3.89 | Ref. 7 |
| PDPP-$g_3$2T$_{0.3}$ | -4.97 | -3.7 | Ref. 8 |
| P($g_4$2T-T) | -4.7 | -3.0 | Ref. 9 |
| PDTDE12 | -4.83 | - | Ref. 10 |
| PBBSe-H | -5.16 | -3.45 | Ref. 11 |
| PF1 | -5.52 | -3.71 | Ref. 12 |
| PM6 | -5.47 | -3.36 | Ref. 13 |
| PTZ1 | -5.31 | -3.34 | Ref. 14 |
| D18 | -5.51 | -2.77 | Ref. 15 |
| PCPDTBT | -5.30 | -3.6 | Ref. 16 |
| $FeCl_3$ | | -5.8 | Ref. 17,18 |
| F4TCNQ | - | -5.24 | Ref. 19 |

S2-Thermoelectric characteristics of polymers studied in this paper

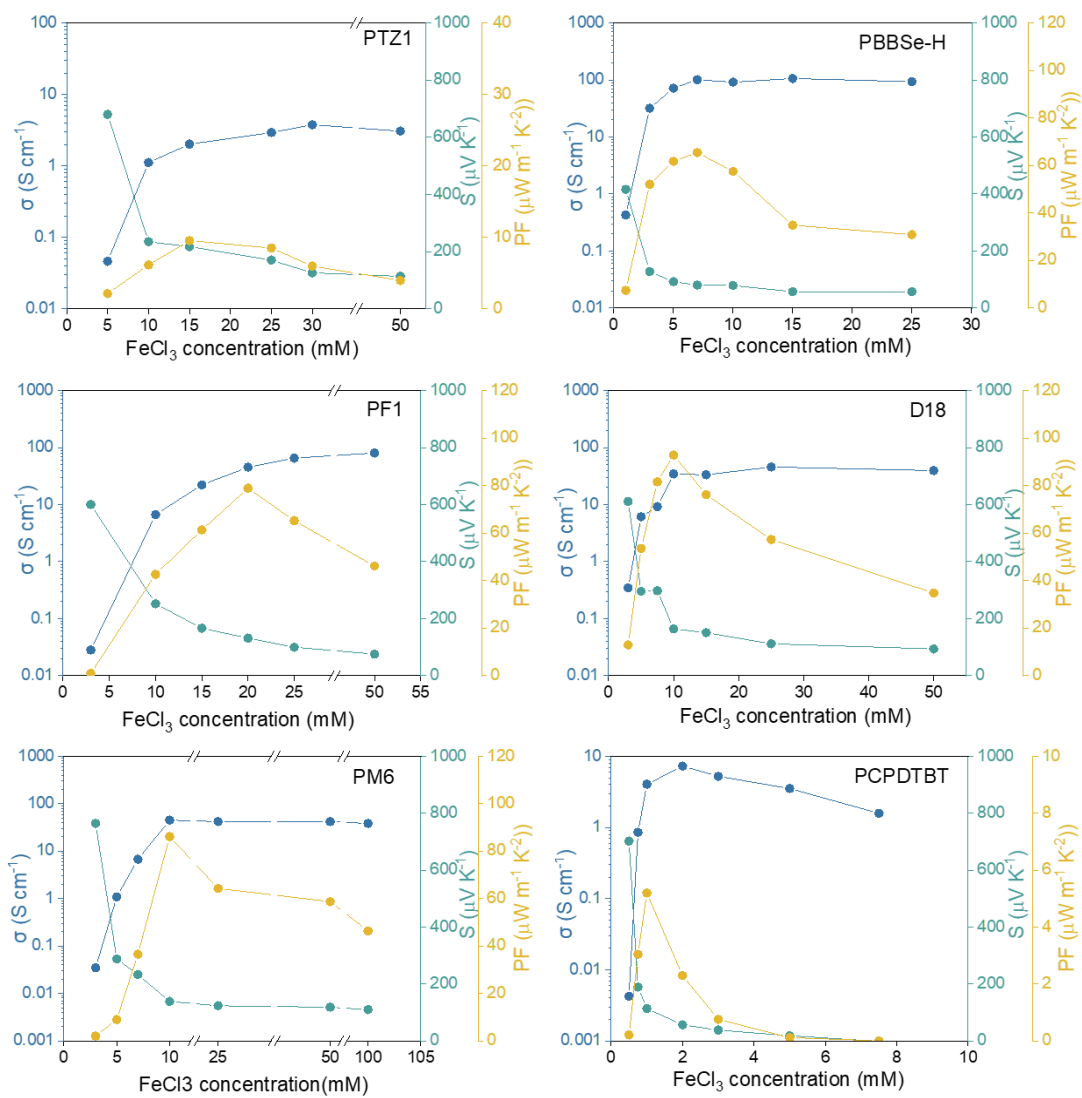

**Figure S2** | Thermoelectric characteristics of all used polymers as a function of FeCl$_3$ concentration.

**Table S2** The optimal thermoelectric properties for individual polymers corresponding to Figure S2.

| Materials | Dopant | $\sigma$ (S m$^{-1}$) | S ($\mu$V K$^{-1}$) | PF ($\mu$W m$^{-1}$K$^{-2}$) | Reference |
|---|---|---|---|---|---|
| P3HT | Fe (III) triflimide | 22958.36 | 36 | 30.51 | Ref. 2 |
| PSBTBT | Fe (III) triflimide | 1854.78 | 44 | 3.66 | Ref. 2 |
| PDPP3T | FeCl$_3$ | 5437.35 | 217 | 256.53 | Ref. 20 |
| PEDOT: PSS | - | 2459.61 | 93 | 21.29 | Ref. 21 |
| IDT-BT | FeCl$_3$ | 637.69 | 331 | 70.15 | Ref. 22 |
| PBTTT | BCF | 4180.66 | 179 | 135.33 | Ref. 23 |
| DPP-BTz | FeCl$_3$ | 49537.84 | 49 | 123.10 | Ref. 24 |
| PDPP-Se | FeCl$_3$ | 89881.13 | 57 | 292.02 | Ref. 7 |
| PDPP-g$_3$2T$_{0.3}$ | FeCl$_3$ | 36156.78 | 55 | 109.37 | Ref. 8 |
| P(g$_4$2T-T) | F4TCNQ | 9767.69 | 26 | 6.57 | Ref. 25 |
| PDTDE12 | F4TCNQ | 11847.86 | 29 | 9.96 | Ref. 10 |
| PBBSe-H | FeCl$_3$ | 10097.18 | 80 | 65.45 | **This work** |
| PF1 | FeCl$_3$ | 4528.78 | 132 | 78.91 | **This work** |
| PM6 | FeCl$_3$ | 5804.06 | 123 | 88.52 | **This work** |
| PTZ1 | FeCl$_3$ | 201.35 | 218 | 9.54 | **This work** |
| D18 | FeCl$_3$ | 3450.95 | 164 | 92.83 | **This work** |
| PCPDTBT | FeCl$_3$ | 404.33 | 113 | 51.95 | **This work** |

S3-Thermoelectric characteristics for S–σ curves of polymers studied in this paper

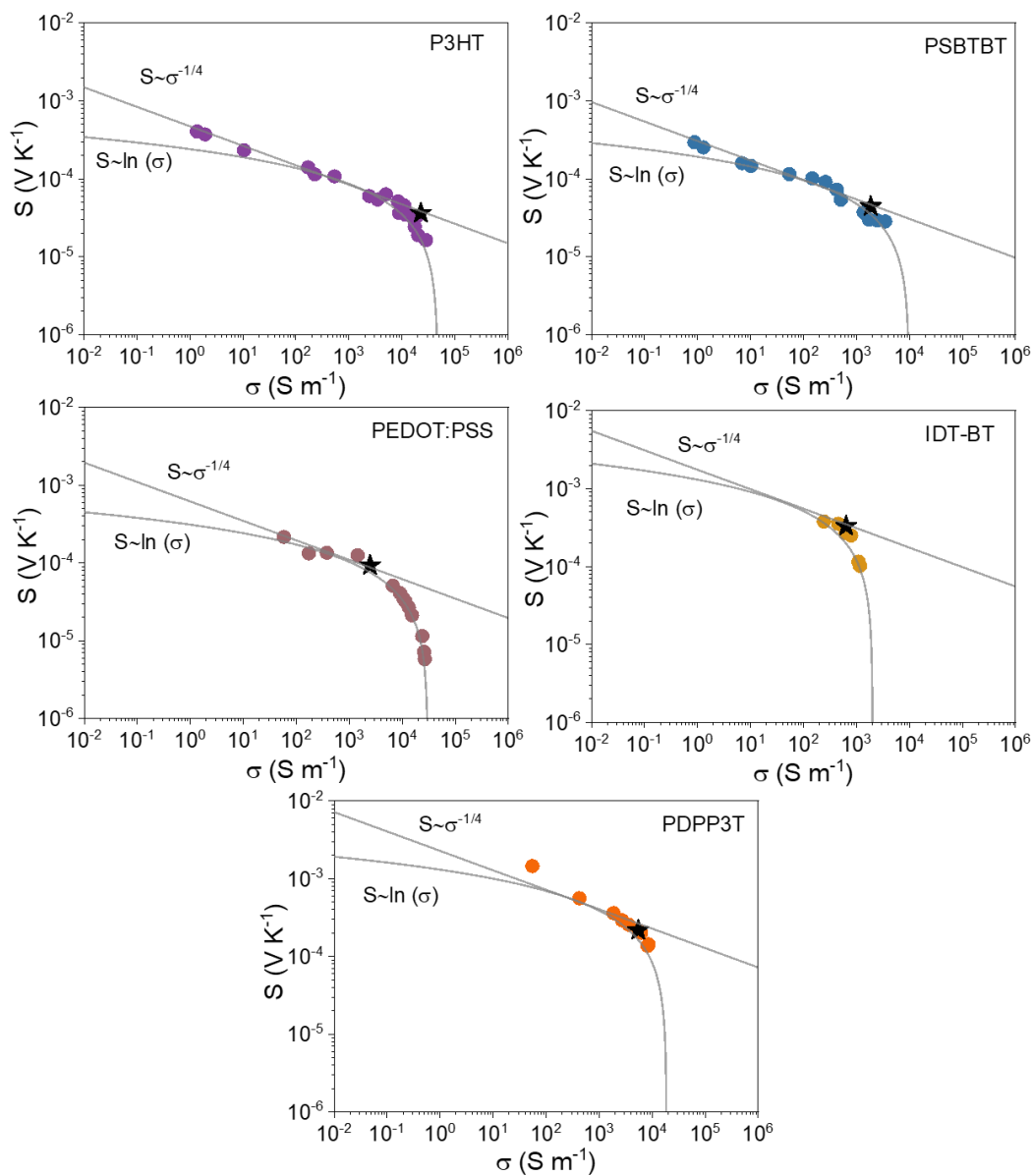

**Figure S3 |** *S-σ* relations of polymers from literature studied in this paper with analytical model with $S \propto \sigma^{-0.25}$ and $S \propto \ln(\sigma)$. The star symbol represents the maximum PF ($PF_{max}$)

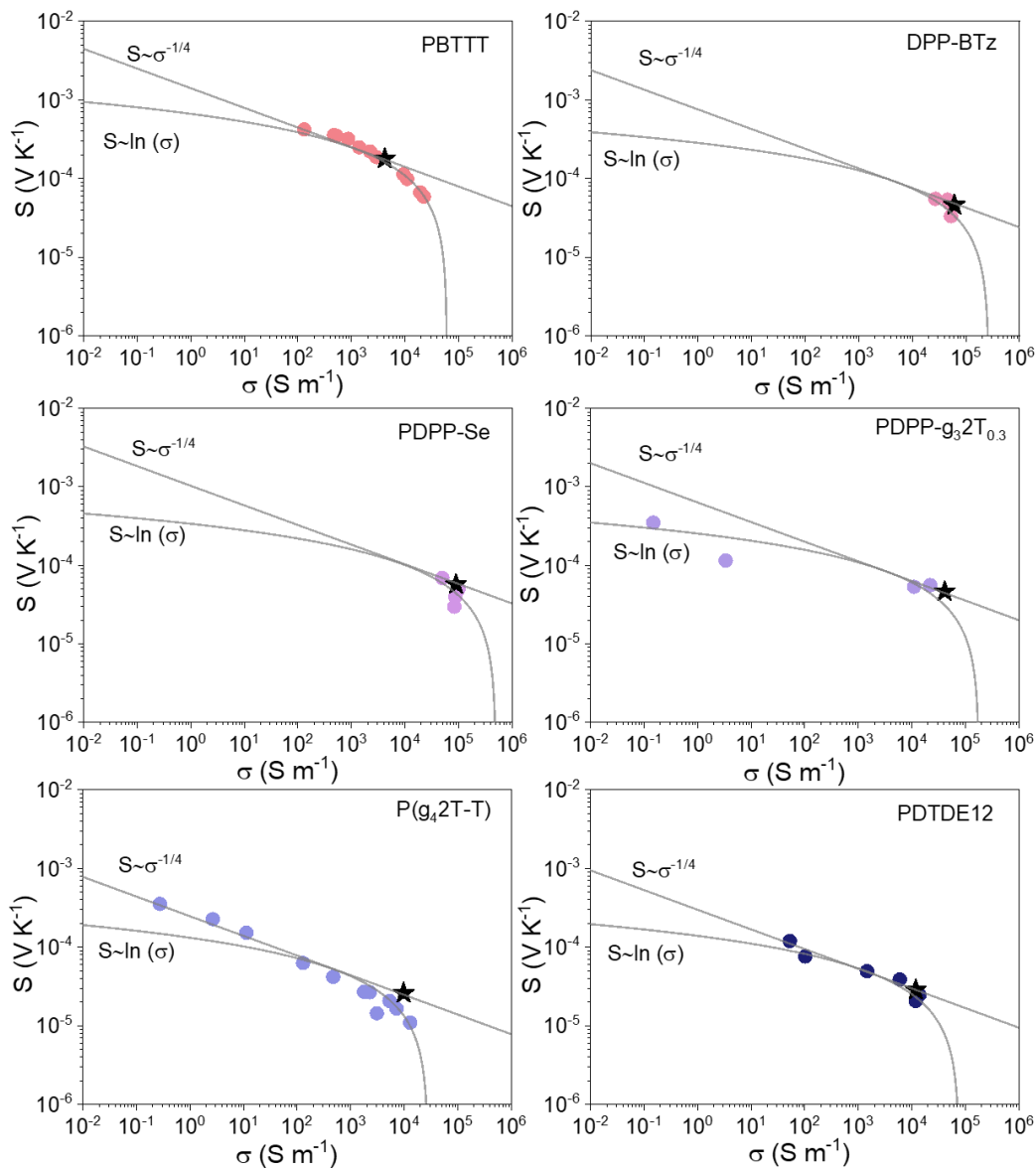

**Figure S4** | *S*-*σ* relations of polymers from literature used in this paper with analytical model with $S \propto \sigma^{-0.25}$ and $S \propto \ln(\sigma)$. The star symbol represents the maximum PF (*PF$_{max}$*)

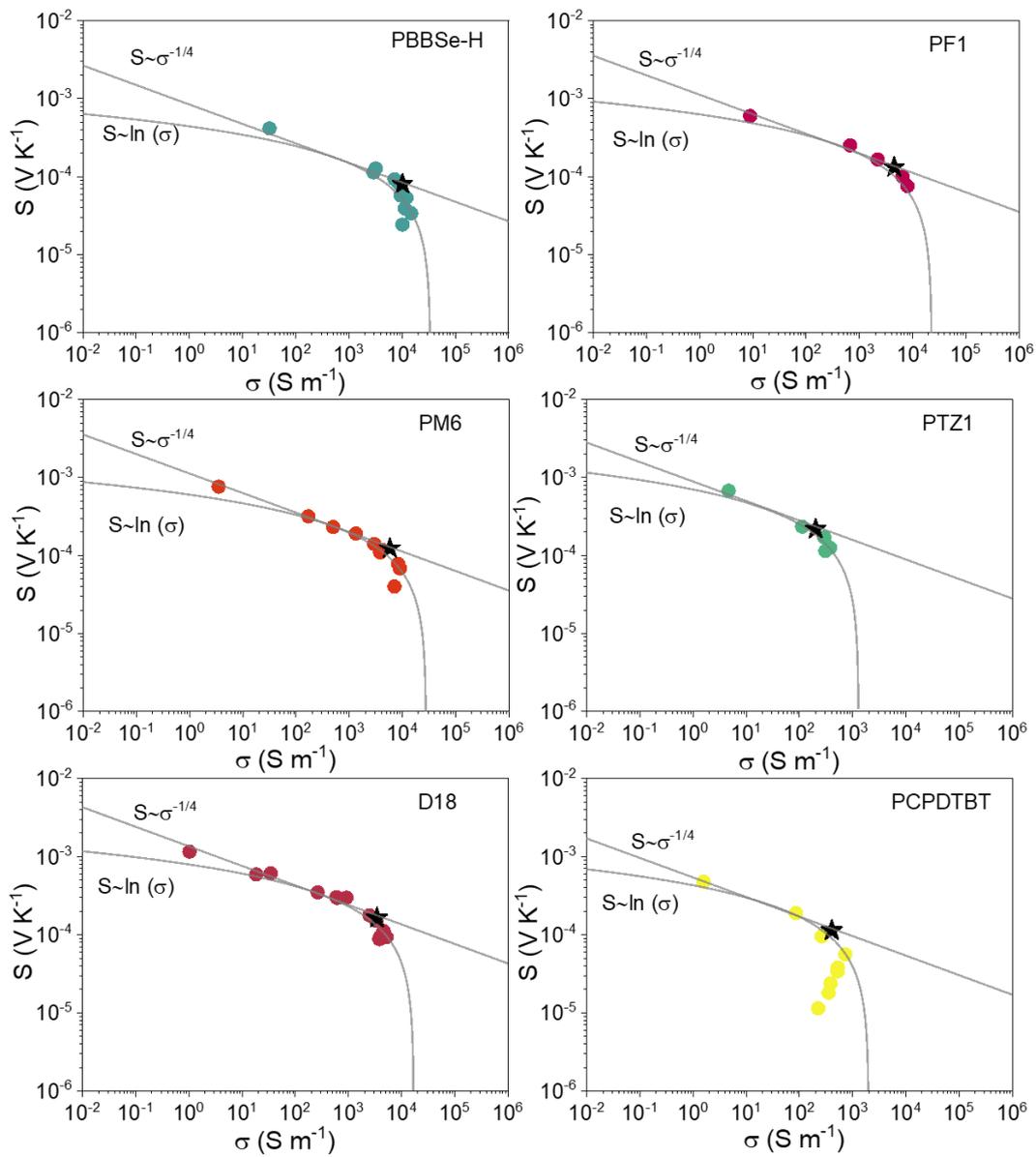

**Figure S5** | $S$-$\sigma$ relations of polymers from our work studied in this paper with analytical model with $S \propto \sigma^{-0.25}$ and $S \propto \ln(\sigma)$. The star symbol represents the maximum PF ($PF_{max}$).

**Table S3** Fitting parameters of the $S$–$\sigma$ curves for the polymers studied in this paper, corresponding to Figure S3, Figure S4 and Figure S5. For the power law fitting $S \propto \sigma^{-0.25}$ the equation used is $S = \frac{k_B}{e}\left(\frac{\sigma}{\sigma_\alpha}\right)^{-0.25}$. For the logarithmic fitting $S \propto \ln(\sigma)$, the equation used is $S = -\frac{k_B}{e}\frac{1}{\beta}\ln\left(\frac{\sigma_{max}}{\sigma}\right)$, where $k_B$ is the Boltzmann constant, $e$ is the elementary charge.

| Materials | $S \propto \sigma^{-0.25}$ | $S \propto \ln(\sigma)$ | |
| --- | --- | --- | --- |
| | $\sigma_\alpha$ | $\beta$ | $\sigma_{max}$ |
| P3HT | 876.52 | 3.9058 | 49000 |
| PSBTBT | 155.22 | 4.1739 | 10000 |
| PDPP3T | 474258.81 | 0.6566 | 19000 |
| PEDOT: PSS | 2559.04 | 2.8979 | 31000 |
| IDT-BT | 166188.86 | 0.5078 | 2050 |
| PBTTT | 68897.59 | 1.44 | 61000 |
| DPP-BTz | 5782.27 | 3.8231 | 266400 |
| PDPP-Se | 19887.94 | 3.3701 | 496450 |
| PDPP-g$_3$2T$_{0.3}$ | 2813.23 | 4.1387 | 180000 |
| P(g$_4$2T-T) | 65.16 | 6.7698 | 28000 |
| PDTDE12 | 139.12 | 7.0387 | 77000 |
| PBBSe-H | 9050 | 2.0455 | 34000 |
| PF1 | 28091.08 | 1.3852 | 23000 |
| PM6 | 28091.08 | 1.4758 | 28000 |
| PTZ1 | 11013.94 | 0.8852 | 1300 |
| D18 | 59125.16 | 1.0658 | 17000 |
| PCPDTBT | 1493.12 | 1.5387 | 2000 |

## S4-Temperature dependence of conductivity and model fittings

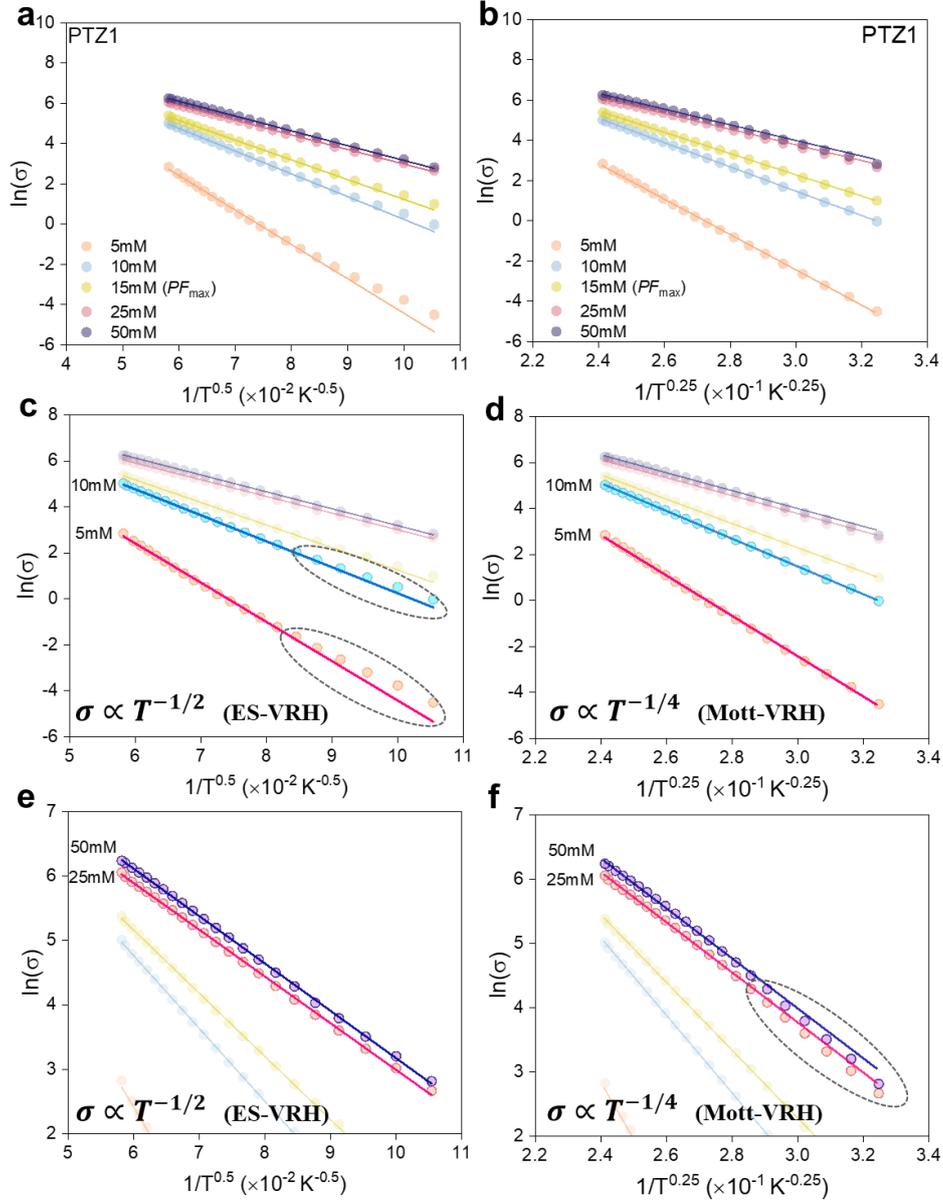

**Figure S6 |** Logarithmic experimental conductivity of PTZ1 as a function of temperature at varying FeCl$_3$ dopant concentrations. a, c, e) The corresponding temperature-dependent conductivities fitted using the Efros-Shklovskii variable-range hopping (ES-VRH) model (ln($\sigma$) $\propto T^{-0.5}$). b, d, f) The corresponding temperature-dependent conductivities fitted using the Mott-type variable range hopping (Mott-VRH) model (ln($\sigma$) $\propto T^{-0.25}$). The results indicate that at low doping concentrations—up to the point of maximum power factor—the conductivity follows the Mott-VRH behavior, while a transition to ES-VRH behavior is observed at and beyond the maximum power factor. Dashed circles highlight regions with noticeable deviations between the experimental data and the respective model fits.

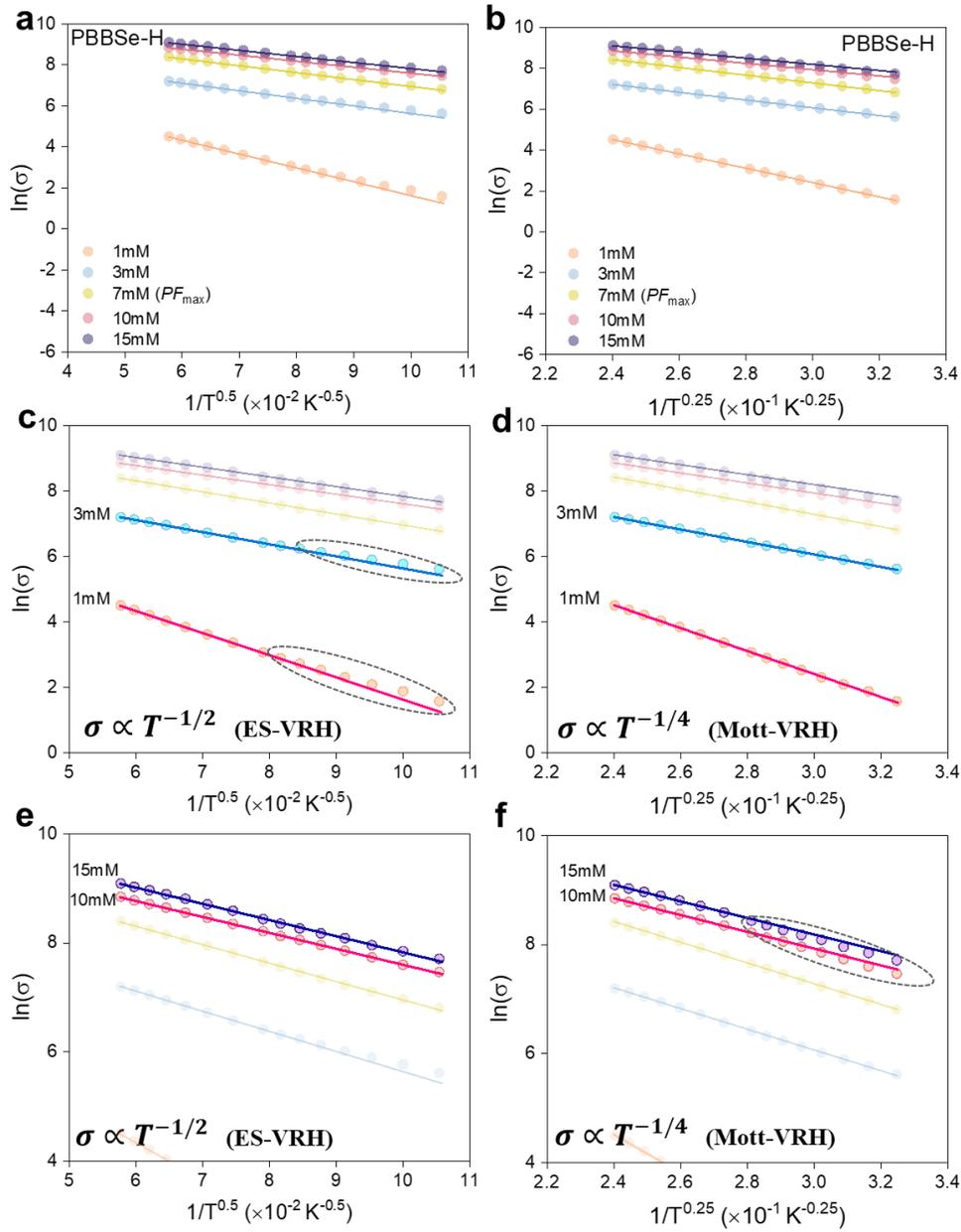

**Figure S7** | Logarithmic experimental conductivity of PBBSe-H as a function of temperature at varying FeCl$_3$ dopant concentrations. a, c, e) The corresponding temperature-dependent conductivities fitted using the Efros-Shklovskii variable-range hopping (ES-VRH) model (ln($\sigma$) $\propto T^{-0.5}$). b, d, f) The corresponding temperature-dependent conductivities fitted using the Mott-type variable range hopping (Mott-VRH) model (ln($\sigma$) $\propto T^{-0.25}$). The results indicate that at low doping concentrations—up to the point of maximum power factor—the conductivity follows the Mott-VRH behavior, while a transition to ES-VRH behavior is observed at and beyond the maximum power factor. Dashed circles highlight regions with noticeable deviations between the experimental data and the respective model fits.

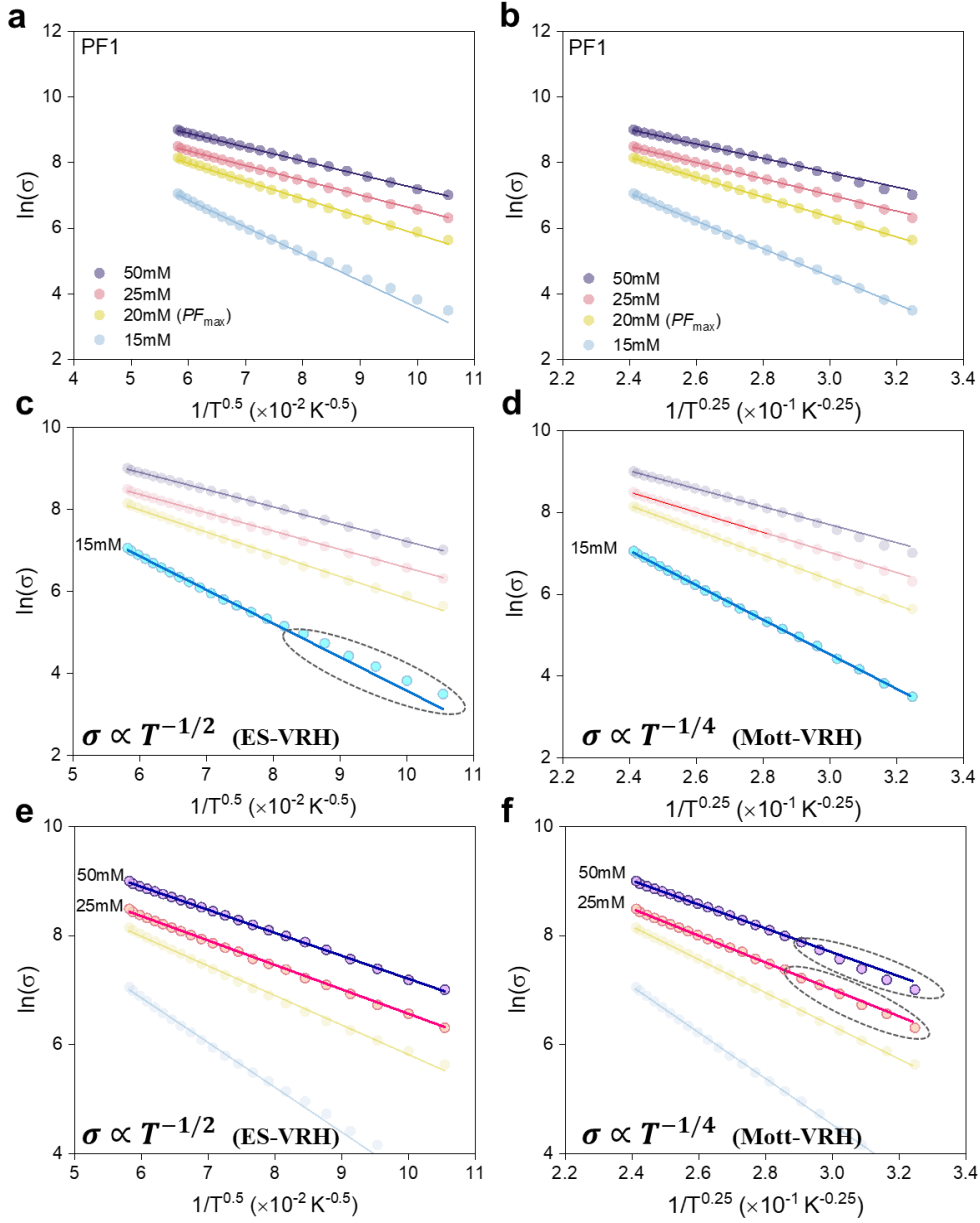

**Figure S8.** Logarithmic experimental conductivity of PF1 as a function of temperature at varying FeCl$_3$ dopant concentrations. a, c, e) The corresponding temperature-dependent conductivities fitted using the Efros-Shklovskii variable-range hopping (ES-VRH) model ($\ln(\sigma) \propto T^{-0.5}$). b, d, f) The corresponding temperature-dependent conductivities fitted using the Mott-type variable range hopping (Mott-VRH) model ($\ln(\sigma) \propto T^{-0.25}$). The results indicate that at low doping concentrations—up to the point of maximum power factor—the conductivity follows the Mott-VRH behavior, while a transition to ES-VRH behavior is observed at and beyond the maximum power factor. Dashed circles highlight regions with noticeable deviations between the experimental data and the respective model fits.

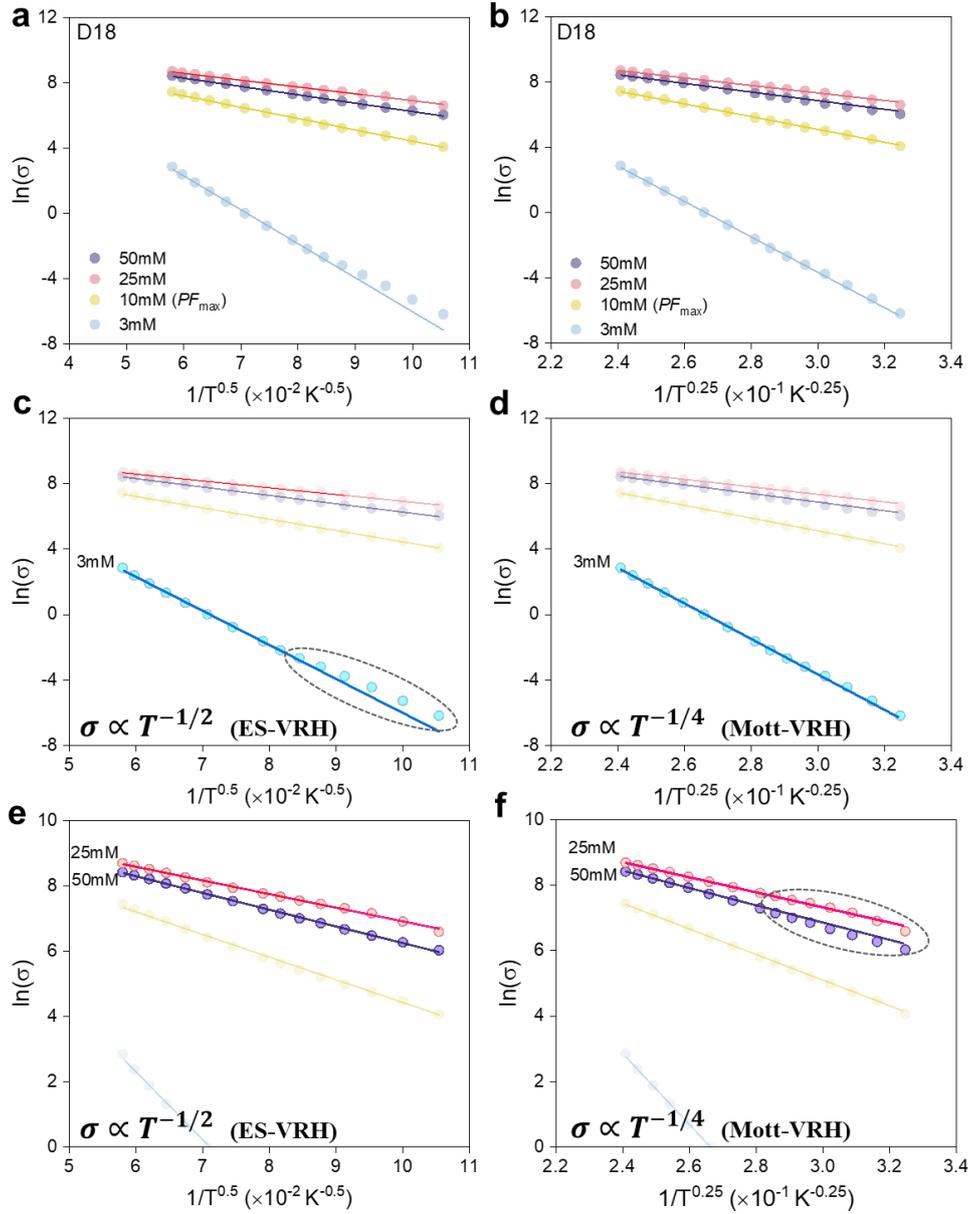

**Figure S9.** Logarithmic experimental conductivity of D18 as a function of temperature at varying FeCl$_3$ dopant concentrations. a, c, e) The corresponding temperature-dependent conductivities fitted using the Efros-Shklovskii variable-range hopping (ES-VRH) model ($\ln(\sigma) \propto T^{-0.5}$). b, d, f) The corresponding temperature-dependent conductivities fitted using the Mott-type variable range hopping (Mott-VRH) model ($\ln(\sigma) \propto T^{-0.25}$). The results indicate that at low doping concentrations—up to the point of maximum power factor—the conductivity follows the Mott-VRH behavior, while a transition to ES-VRH behavior is observed at and beyond the maximum power factor. Dashed circles highlight regions with noticeable deviations between the experimental data and the respective model fits.

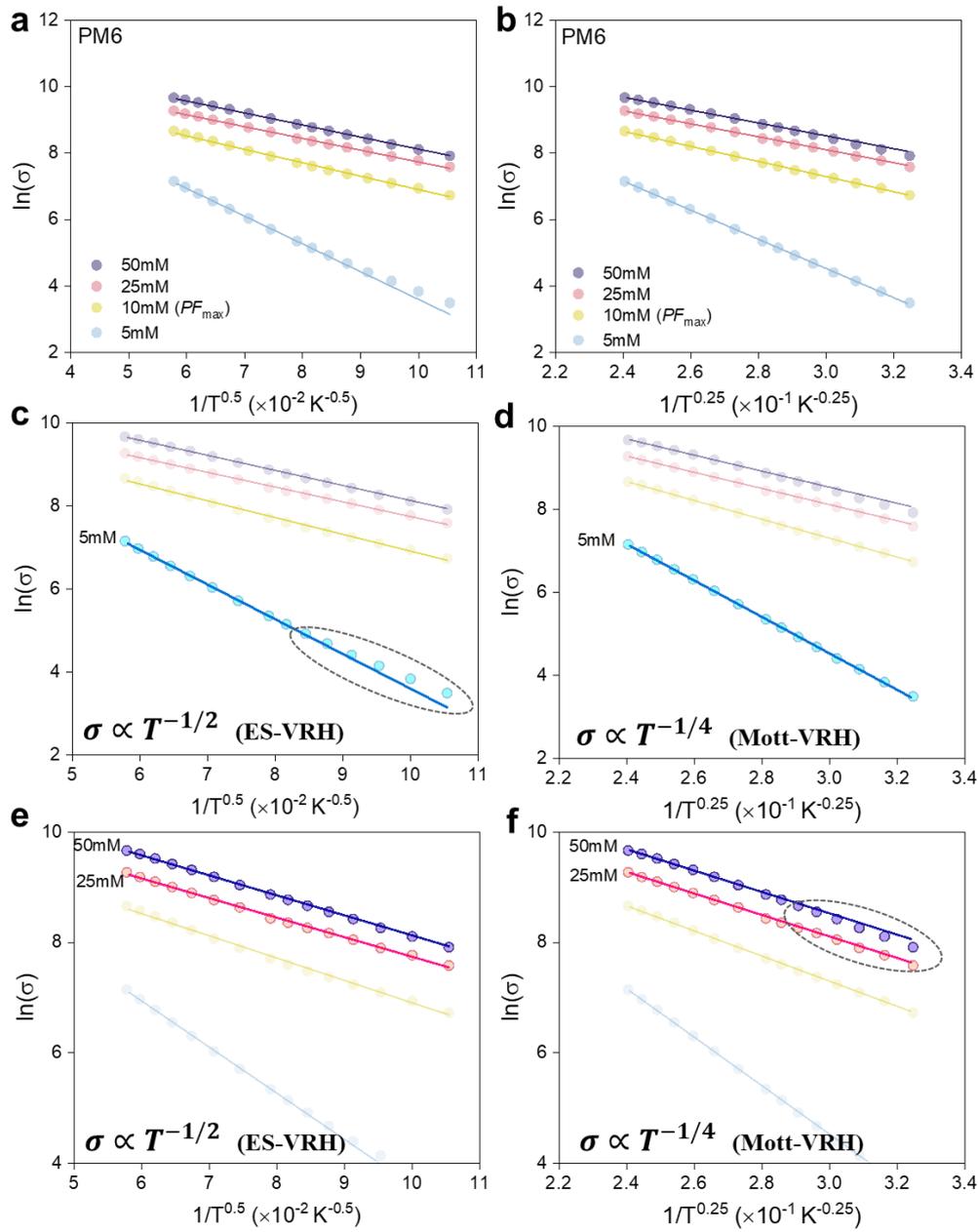

**Figure S10** | Logarithmic experimental conductivity of PM6 as a function of temperature at varying FeCl$_3$ dopant concentrations. a, c, e) The corresponding temperature-dependent conductivities fitted using the Efros-Shklovskii variable-range hopping (ES-VRH) model (ln($\sigma$) $\propto T^{-0.5}$). b, d, f) The corresponding temperature-dependent conductivities fitted using the Mott-type variable range hopping (Mott-VRH) model (ln($\sigma$) $\propto T^{-0.25}$). The results indicate that at low doping concentrations—up to the point of maximum power factor—the conductivity follows the Mott-VRH behavior, while a transition to ES-VRH behavior is observed at and beyond the maximum power factor. Dashed circles highlight regions with noticeable deviations between the experimental data and the respective model fits.

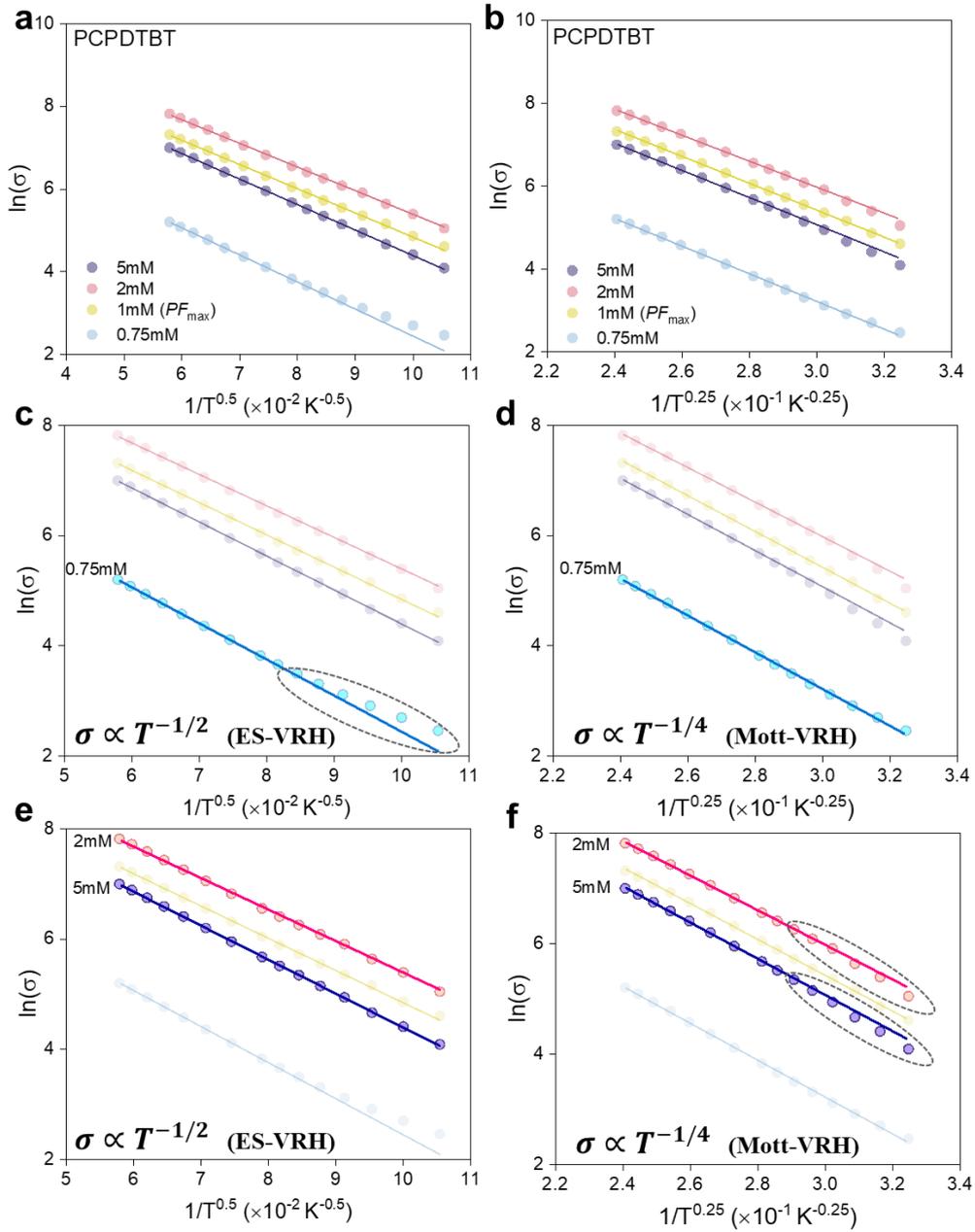

**Figure S11** | Logarithmic experimental conductivity of PCPDTBT as a function of temperature at varying FeCl$_3$ dopant concentrations. a, c, e) The corresponding temperature-dependent conductivities fitted using the Efros-Shklovskii variable-range hopping (ES-VRH) model ($\ln(\sigma) \propto T^{-0.5}$). b, d, f) The corresponding temperature-dependent conductivities fitted using the Mott-type variable range hopping (Mott-VRH) model ($\ln(\sigma) \propto T^{-0.25}$). The results indicate that at low doping concentrations—up to the point of maximum power factor—the conductivity follows the Mott-VRH behavior, while a transition to ES-VRH behavior is observed at and beyond the maximum power factor. Dashed circles highlight regions with noticeable deviations between the experimental data and the respective model fits.

S5-The R-square of the fit to conductivity vs. temperature

Although the temperature-dependent conductivity fits using the Mott-VRH and ES-VRH models yield well-defined exponents, as shown in Figure S6-S10, those clearly indicate a transition from Mott-VRH behavior to ES-VRH behavior at and beyond the point of maximum power factor ($PF_{max}$). In addition to directly fitting the conductivity data with fixed exponents for Mott-VRH and ES-VRH models, we employed a more flexible approach to determine the most suitable hopping exponent $\alpha$.[26] Specifically, we performed systematic fits to the generalized VRH equation:

$$\ln(\sigma) = \ln(\sigma_0) - \left(\frac{T_0}{T}\right)^\alpha$$

by varying α over a range of values (e.g., 0. to 1.0) while keeping $\sigma_0$ and $T_0$ as free parameters. The coefficient of determination (R-square) was then plotted as a function of $\alpha$. This method allows for a more quantitative identification of the optimal hopping regime and reveals a clear transition from $\alpha$ ~ 0.25 (Mott-VRH) at low doping levels to $\alpha$ ~ 0.5 (ES-VRH) near and beyond the maximum power factor.

As shown in **Figure S12-S17**, the extracted $\alpha$ values exhibit a systematic transition from ~ 0.25 to ~ 0.5 with increasing FeCl$_3$ concentration, providing strong evidence for a crossover from Mott-type to ES-type variable-range hopping transport as the carrier concentration increases and Coulomb interactions become significant. The results not only confirm the crossover from Mott-VRH to ES-VRH but also demonstrate the consistency between experimental data and theoretical transport models.

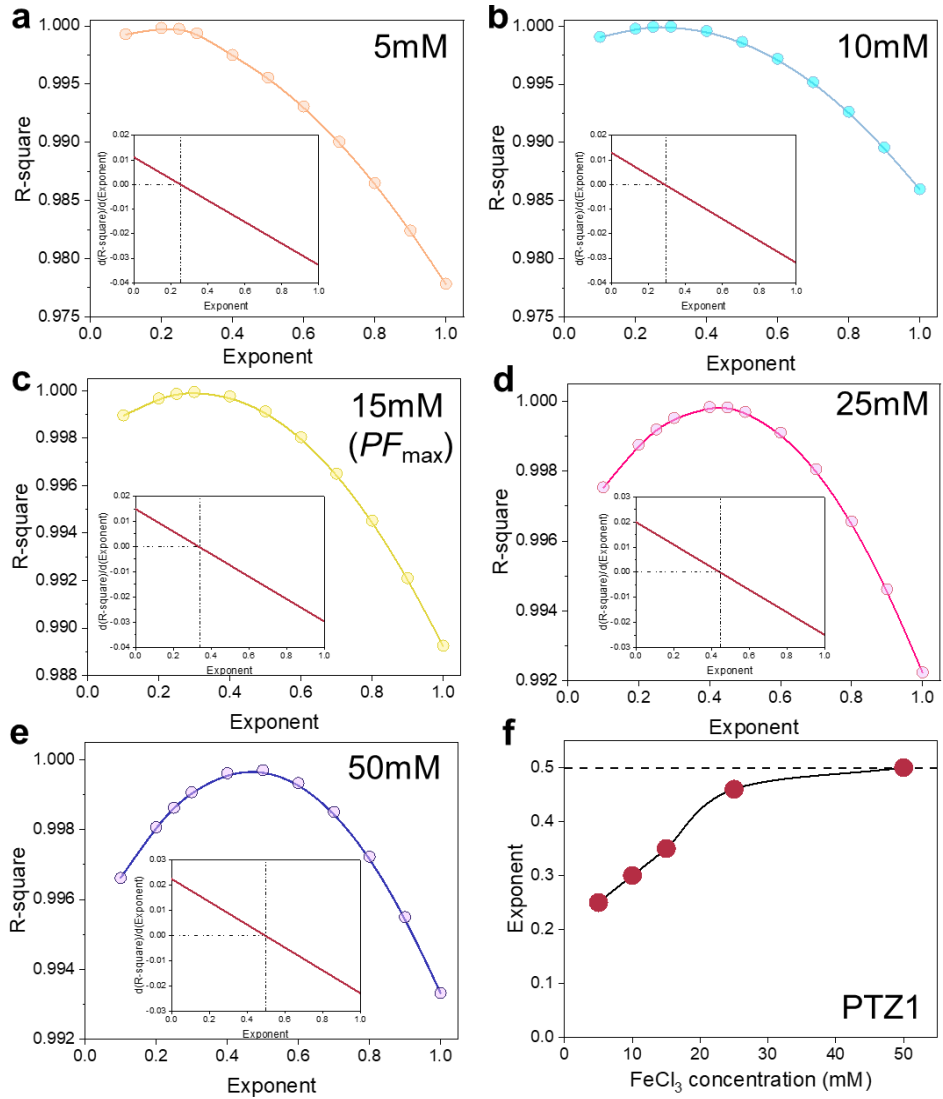

**Figure S12** | The R-square analysis of temperature-dependent conductivity fits for PTZ1 at various FeCl₃ dopant concentrations: a) 5 mM, b) 10 mM, c) 15 mM, d) 25 mM and e) 50 mM; f) Extracted optimal hopping exponent α as a function of FeCl₃ concentration, showing a clear transition from Mott-VRH ($\alpha$ ~ 0.25) to ES-VRH ($\alpha$ ~ 0.5) transport with increasing doping level.

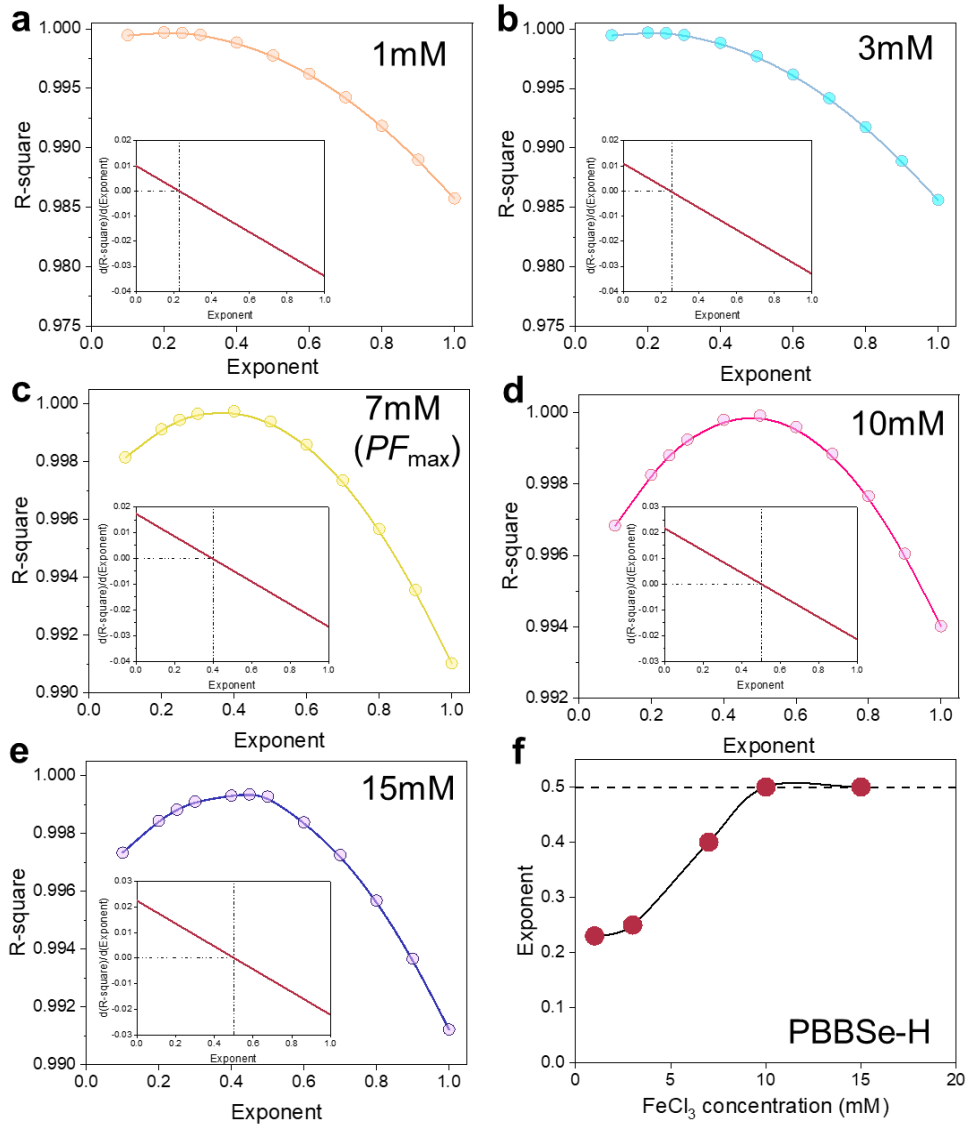

**Figure S13** | The R-square analysis of temperature-dependent conductivity fits for PBBSe-H at various FeCl₃ dopant concentrations: : a) 1 mM, b) 3 mM, c) 7 mM, d) 10 mM and e) 15 mM; f) Extracted optimal hopping exponent α as a function of FeCl₃ concentration, showing a clear transition from Mott-VRH ($\alpha \sim 0.25$) to ES-VRH ($\alpha \sim 0.5$) transport with increasing doping level.

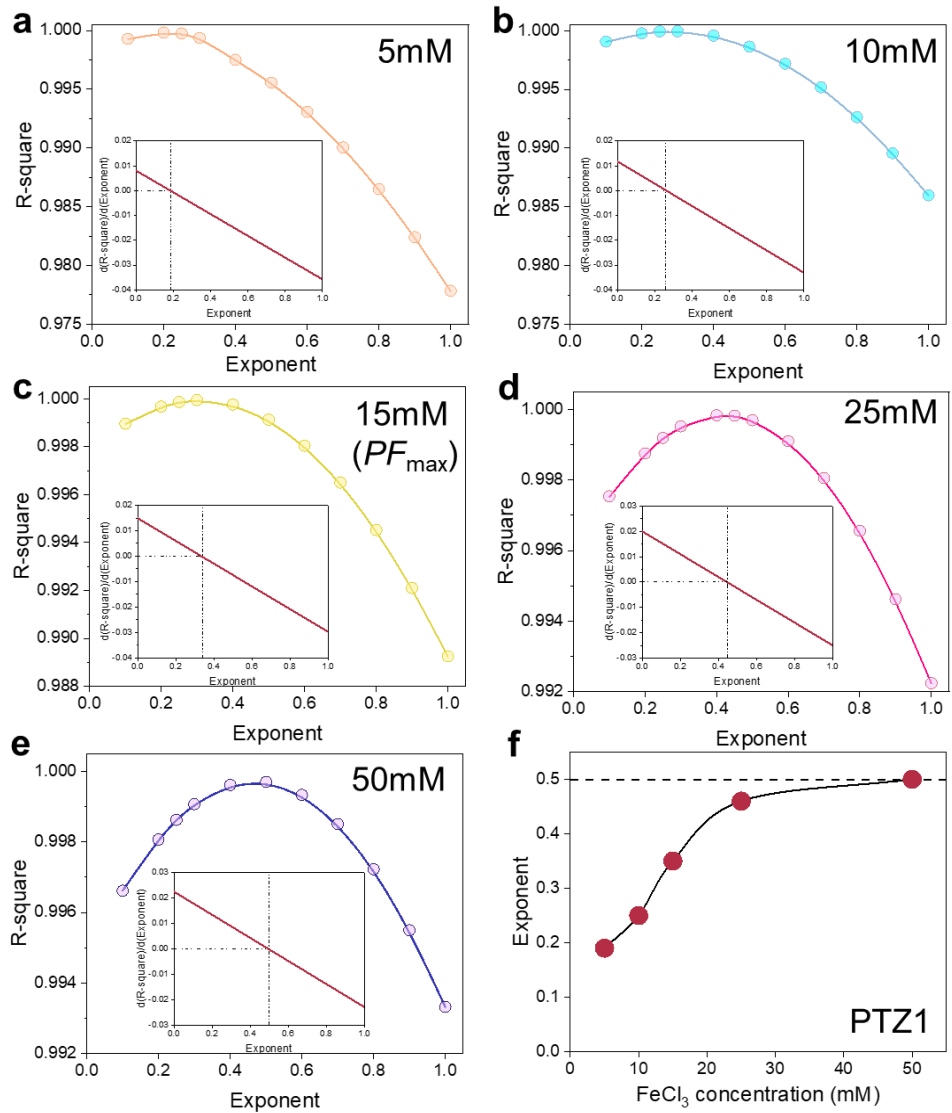

**Figure S14** | The R-square analysis of temperature-dependent conductivity fits for PF1 at various FeCl₃ dopant concentrations: a) 15 mM, b) 20 mM, c) 25 mM and d) 50 mM. f) Extracted optimal hopping exponent α as a function of FeCl₃ concentration, showing a clear transition from Mott-VRH ($\alpha \sim 0.25$) to ES-VRH ($\alpha \sim 0.5$) transport with increasing doping level.

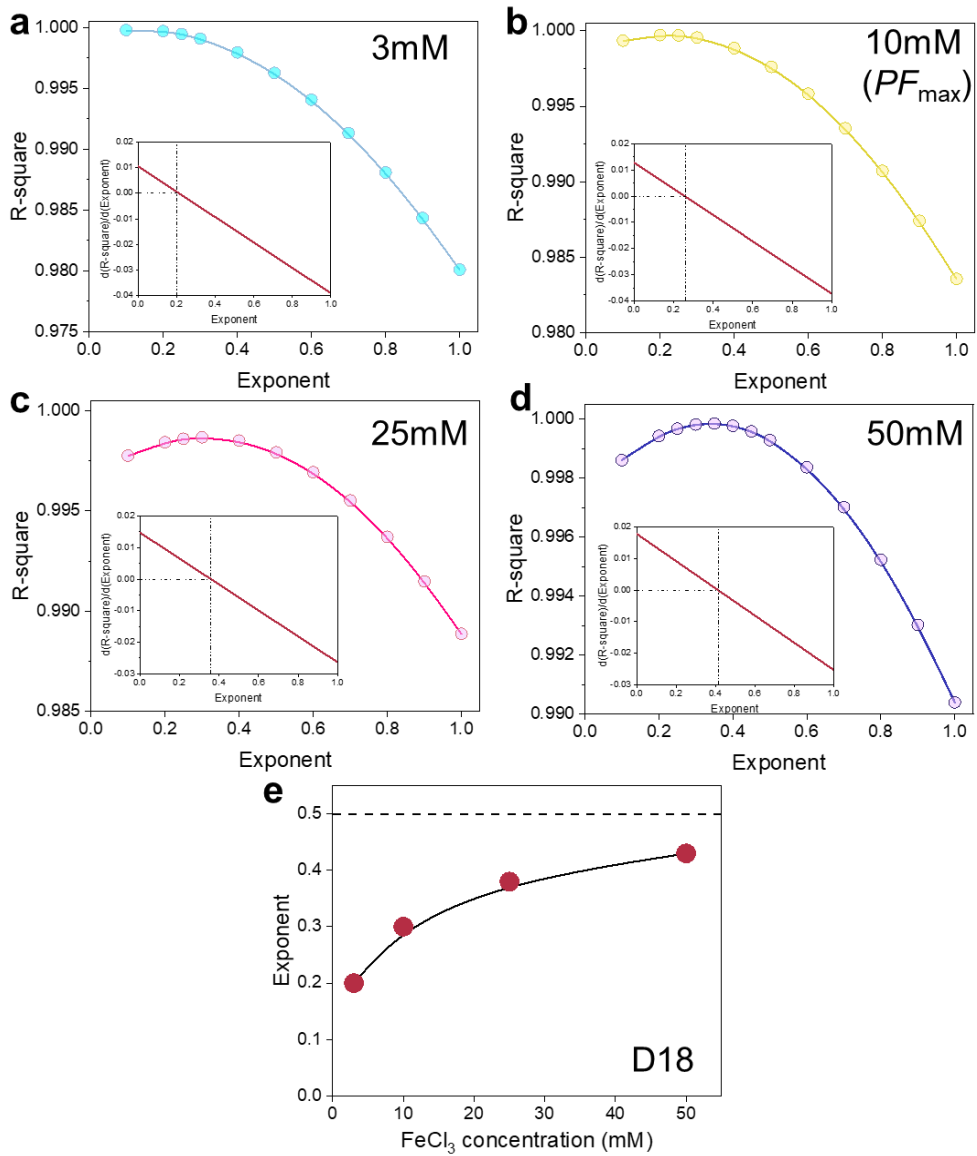

**Figure S15 |** The R-square analysis of temperature-dependent conductivity fits for D18 at various FeCl₃ dopant concentrations: a) 3 mM, b) 10 mM, c) 25 mM and d) 50 mM; f) Extracted optimal hopping exponent $\alpha$ as a function of FeCl₃ concentration, showing a clear transition from Mott-VRH ($\alpha \sim 0.25$) to ES-VRH ($\alpha \sim 0.5$) transport with increasing doping level.

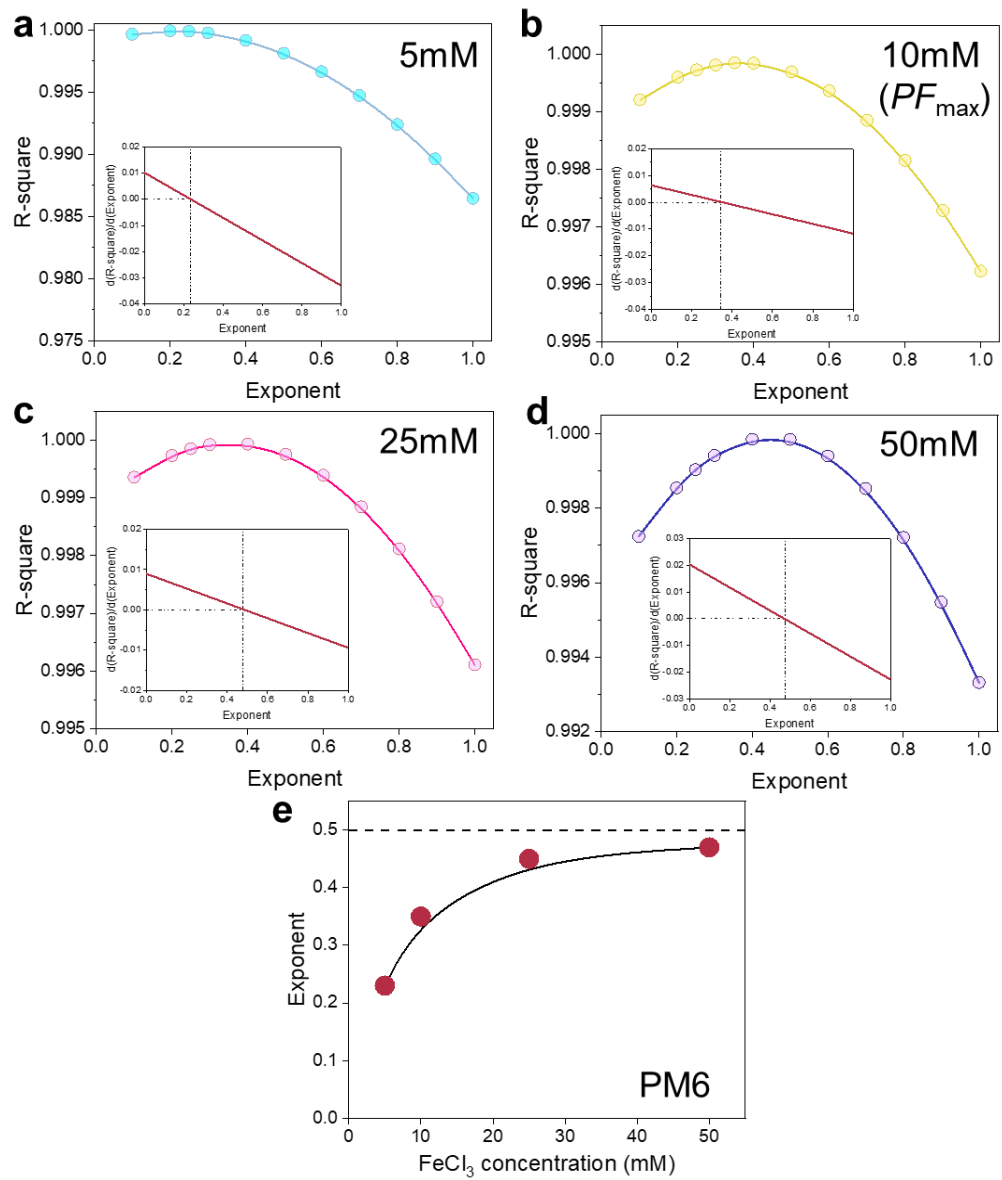

**Figure S16 |** The R-square analysis of temperature-dependent conductivity fits for PM6 at various FeCl₃ dopant concentrations: a) 5 mM, b) 10 mM, c) 25 mM and d) 50 m. f) Extracted optimal hopping exponent α as a function of FeCl₃ concentration, showing a clear transition from Mott-VRH ($\alpha \sim 0.25$) to ES-VRH ($\alpha \sim 0.5$) transport with increasing doping level.

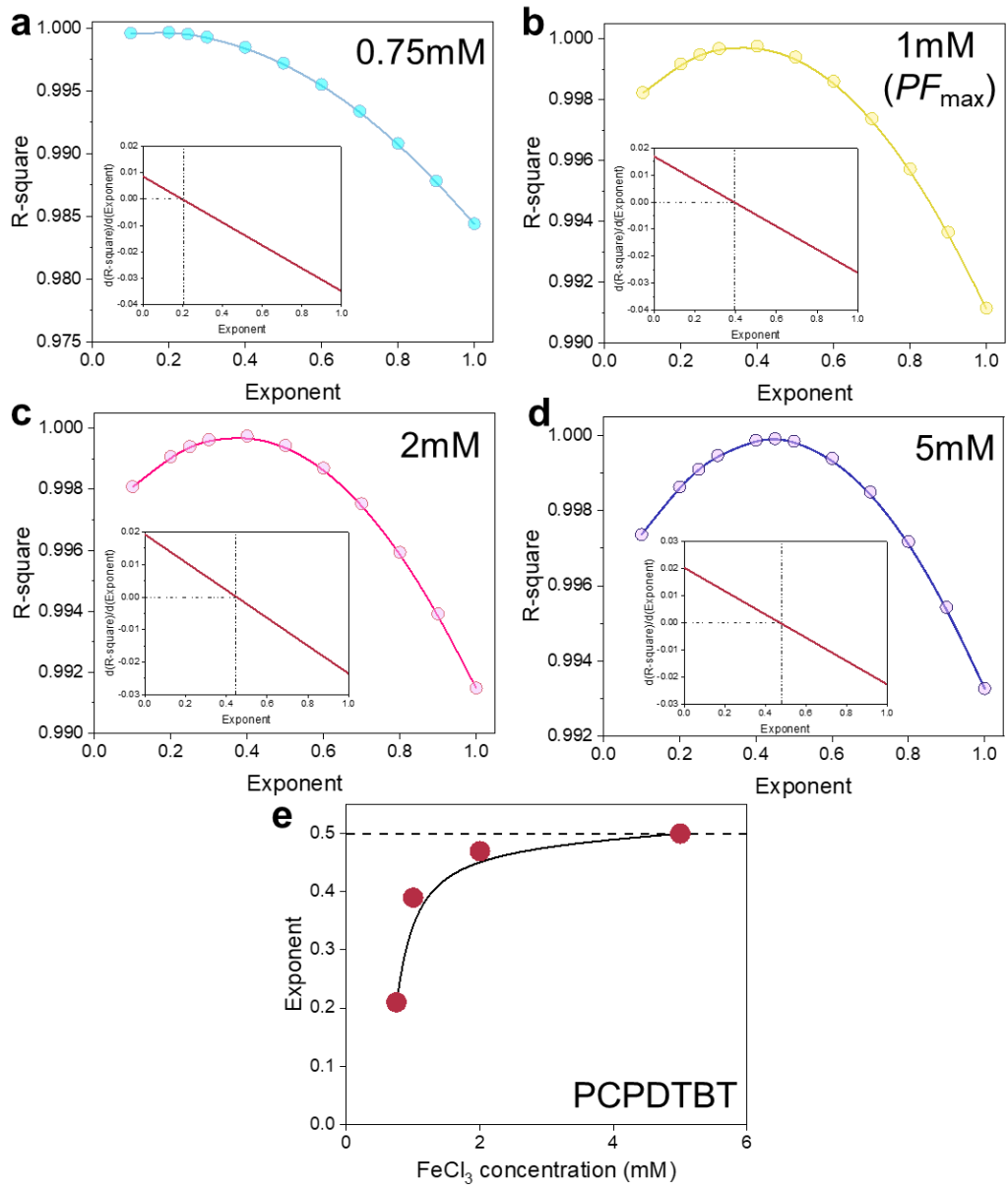

**Figure S17** | The R-square analysis of temperature-dependent conductivity fits for PCPDTBT at various FeCl₃ dopant concentrations: a) 0.75 mM, b) 1 mM, c) 2 mM and d) 5 mM. f) Extracted optimal hopping exponent α as a function of FeCl₃ concentration, showing a clear transition from Mott-VRH ($\alpha \sim 0.25$) to ES-VRH ($\alpha \sim 0.5$) transport with increasing doping level.

## S6-kMC simulations with different parameters

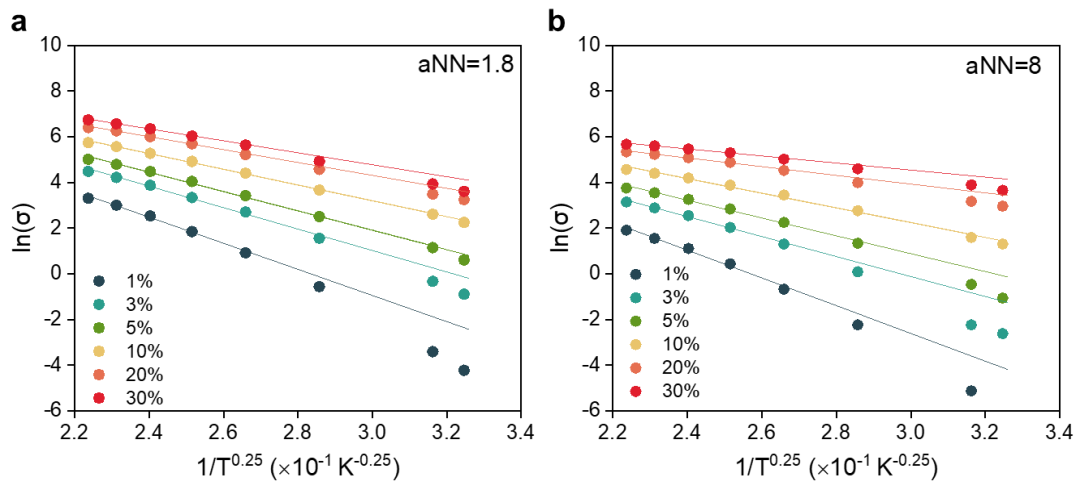

**Figure S18** | Kinetic Monte calculations: a) $a_{NN}$=1.8 nm and b) $a_{NN}$=8 nm. Default kMC parameters: energetic disorder $\sigma_{DOS}$= 75 meV, attempt-to-hop frequency $\nu_0$= $1\times10^{13}$ s$^{-1}$, relative dielectric constant $\varepsilon_r$= 3.6, temperature $T$ from 100 K to 400 K. Same data as in Figure 3c, d of the main text.

S7-The R-square of the fit to conductivity vs. temperature for kMC simulations

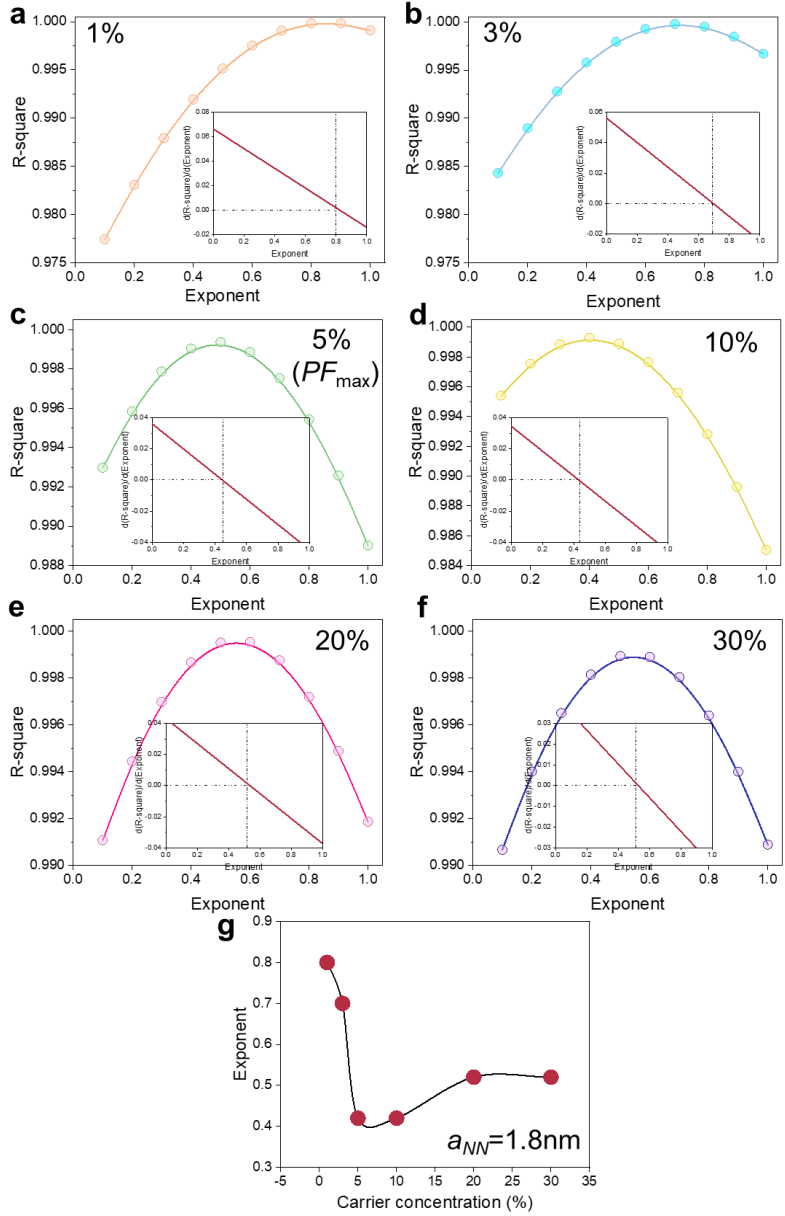

**Figure S19** | The R-square analysis of temperature-dependent conductivity fits for kinetic Monte (kMC) simulations. at various concentrations from 1% to 30% for a-f). R-square as a function of the exponent $\alpha$, obtained by fitting kMC-calculated conductivity over the temperature range of 100-400 K, for carrier concentrations ranging from 1% to 30%. The default kMC parameters: $a_{NN}$=1.8 nm, energetic disorder $\sigma_{DOS}$= 75 meV, attempt-to-hop frequency $\nu_0$= 1×10$^{13}$ s$^{-1}$, relative dielectric constant $\varepsilon_r$= 3.6. g) Simulated optimal $\alpha$ values as a function of carrier concentration, showing a clear transition from Mott-type (around 5-10%) to ES-type hopping behavior (at 20-30%). The apparent upswing at low concentrations (below 5%) is not observed in experiments, which can be attributed to the lack of experimental data at such low doping regimes, which lies outside the experimental scope of this study.

S8-The change of DOS by kMC simulation with different parameters.

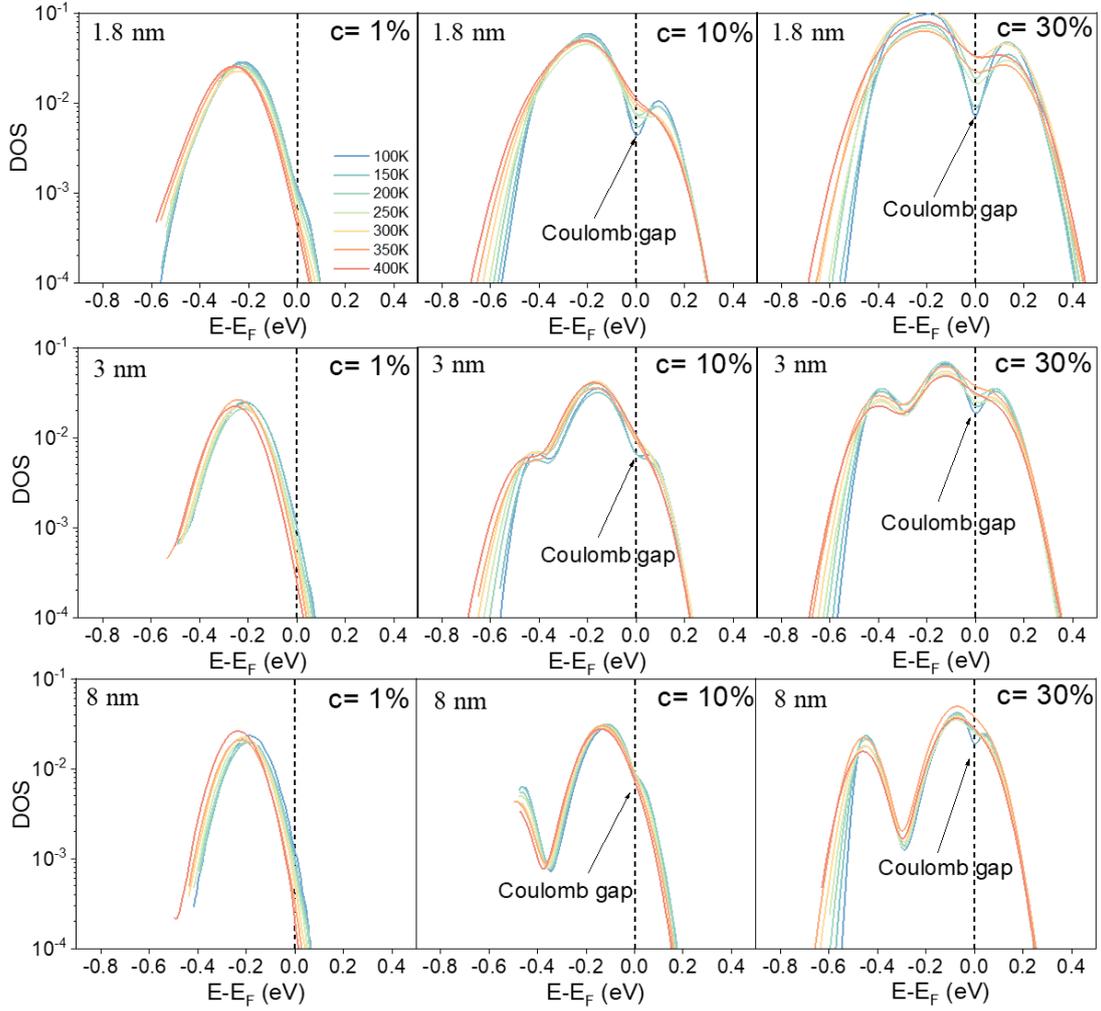

**Figure S20** | Evolution of DOS from kMC simulation at different inter-site distances $a_{NN}$. The default kMC parameters: $a_{NN}$ = 1.8 nm, 3nm, and 8 nm, respectively, energetic disorder $\sigma_{DOS}$ = 75 meV, attempt-to-hop frequency $\nu_0$ = 1×10$^{13}$ s$^{-1}$, relative dielectric constant $\varepsilon_r$ = 3.6, temperature $T$ from 100 K to 400 K. Holes concentration $c$ from 1% to 30%. These results suggest that, at a fixed hole concentration, the formation of a soft Coulomb gap is suppressed with increasing the inter-site distance $a_{NN}$ (and indirectly the localization length which is set as a fraction (0.33) of that), as reflected by larger values of $a_{NN}$. This trend is consistent with the Coulomb interaction equation ($E_C = \frac{e}{4\pi\varepsilon_r\varepsilon_0 a}$), which indicates that a larger spatial separation between carriers reduces the effective Coulomb repulsion, thereby delaying the onset or diminishing the depth of the gap.

Our simulations demonstrate that a sufficiently high doping level is required for Coulomb interactions to induce a pronounced soft Coulomb gap, thereby driving the transition from Mott-VRH to ES-VRH. At even higher doping levels—particularly when the carrier occupation approaches one electron per site—our simulations reveal the emergence of a larger suppression in the DOS below the Fermi energy ($E < E_F$) that cannot be accounted for by the soft ES gap alone. Instead, this feature reflects the formation of a Mott–Hubbard-type hard gap, arising from the energetic penalty of double occupation of localized states. This interpretation is strongly supported by the recent findings of Xu et al.[27], who attributed the inversion of the Seebeck coefficient and drop in electrical conductivity in highly doped BBL to the opening of a hard Coulomb gap caused by strong electron–electron repulsion and multiple occupancy. For example, at c = 30%, our simulations reveal a pronounced suppression of the DOS at $E - E_F > 0$, consistent with the onset of double-occupation blocking.

To interpret these results correctly, it is essential to recall that in our DOS convention, occupied states lie at $E - E_F > 0$, and empty states at $E - E_F < 0$. This perspective enables a clear distinction between the soft Coulomb gap, centered at the Fermi level due to long-range Coulomb interactions, and the hard gap below $E_F$, which reflects on-site repulsion and the formation of localized states resistant to double occupation. Together, these two regimes clarify the evolving interplay between doping level, Coulomb interaction strength, and the resulting charge transport behavior.

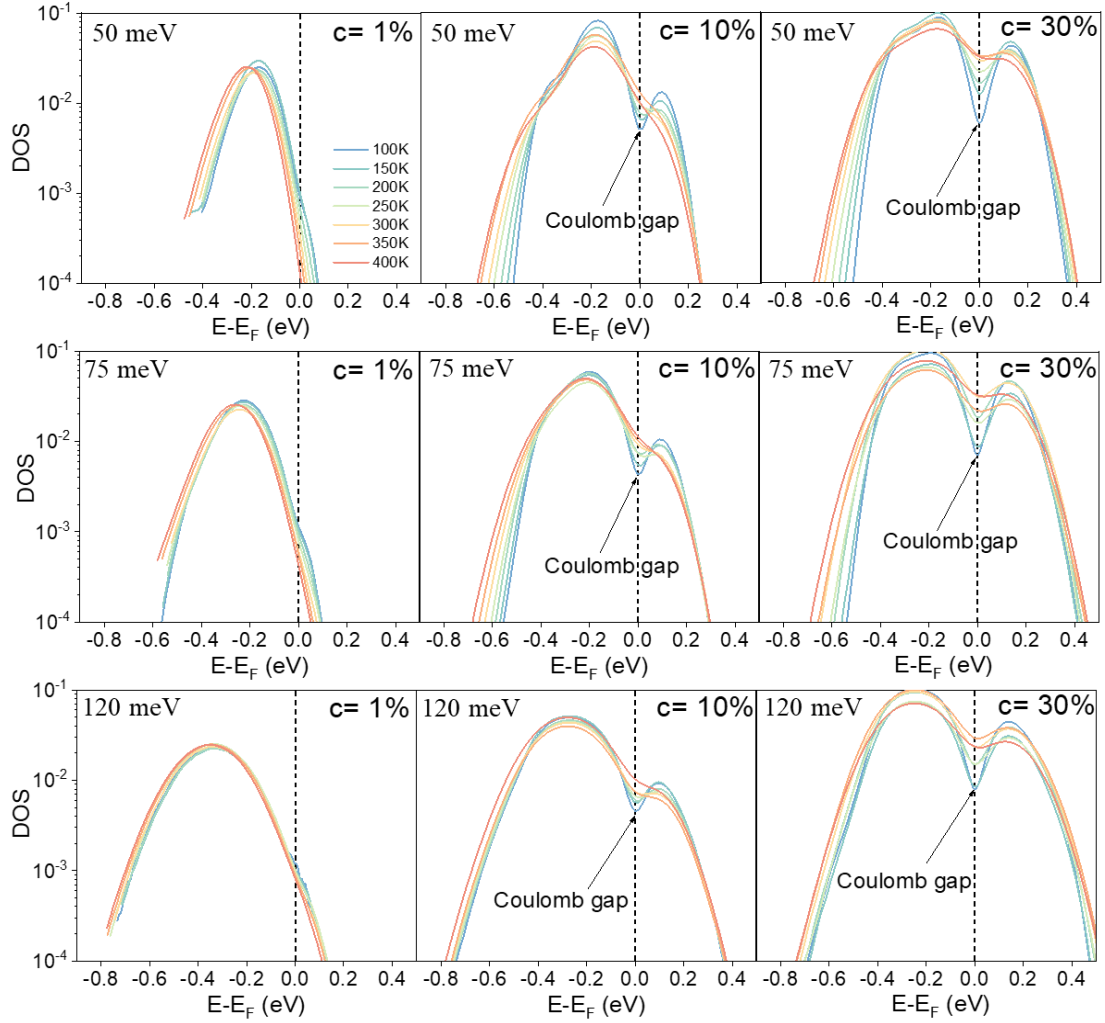

**Figure S21** | Evolution of DOS from kMC simulation at different energetic disorder $\sigma_{DOS}$. The default kMC parameters: $a_{NN}$= 1.8 nm, $\sigma_{DOS}$= 50 meV, 75 meV, and 120 meV, respectively, attempt-to-hop frequency $\nu_0 = 1\times10^{13}\,s^{-1}$, relative dielectric constant $\varepsilon_r$= 3.6, temperature $T$ from 100 K to 400 K. Holes concentration $c$ from 1% to 30%. These results suggest that, at a fixed Hole concentration, the formation of a soft Coulomb gap is only weakly influenced by increasing energetic disorder. However, it is worth noting that a smaller $\sigma_{DOS}$ significantly benefits the conductivity, likely leading to an improved power factor.

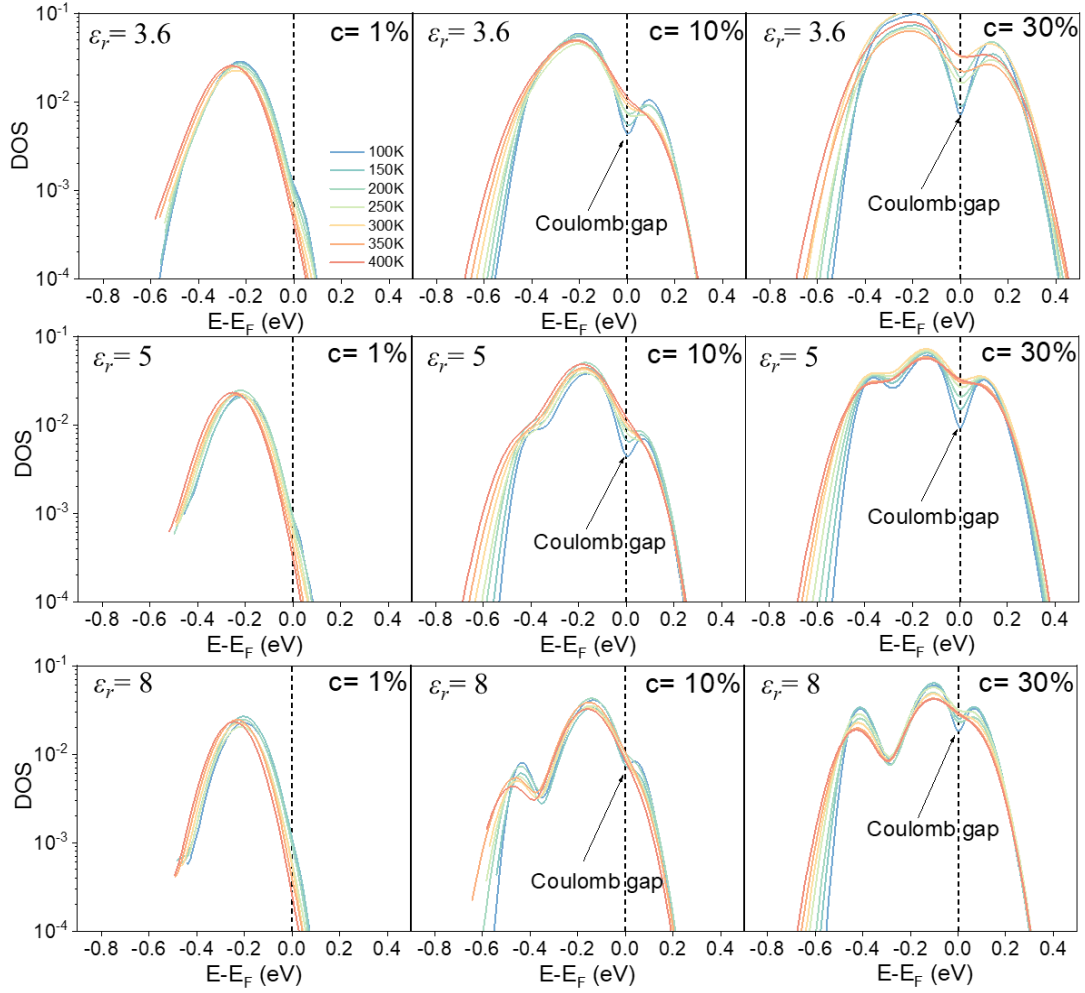

**Figure S22** | Evolution of DOS from kMC simulation at different relative dielectric constant $\varepsilon_r$. The default kMC parameters: $a_{NN}$= 1.8 nm, energetic disorder $\sigma_{DOS}$= 75 meV, attempt-to-hop frequency $\nu_0 = 1\times10^{13}$ s$^{-1}$, $\varepsilon_r$ = 3.6, 5 and 8, respectively, temperature $T$ from 100 K to 400 K. Holes concentration $c$ from 1% to 30%. These results suggest that, at a fixed hole concentration, the formation of a soft Coulomb gap is suppressed with increasing material permittivity, as reflected by larger values of the dielectric constant ($\varepsilon_r$). This trend is consistent with the Coulomb interaction equation ($E_C = \frac{e}{4\pi\varepsilon_r\varepsilon_0 a}$), which shows that higher permittivity reduces the effective Coulomb repulsion, thereby delaying the onset or reducing the depth of the gap.

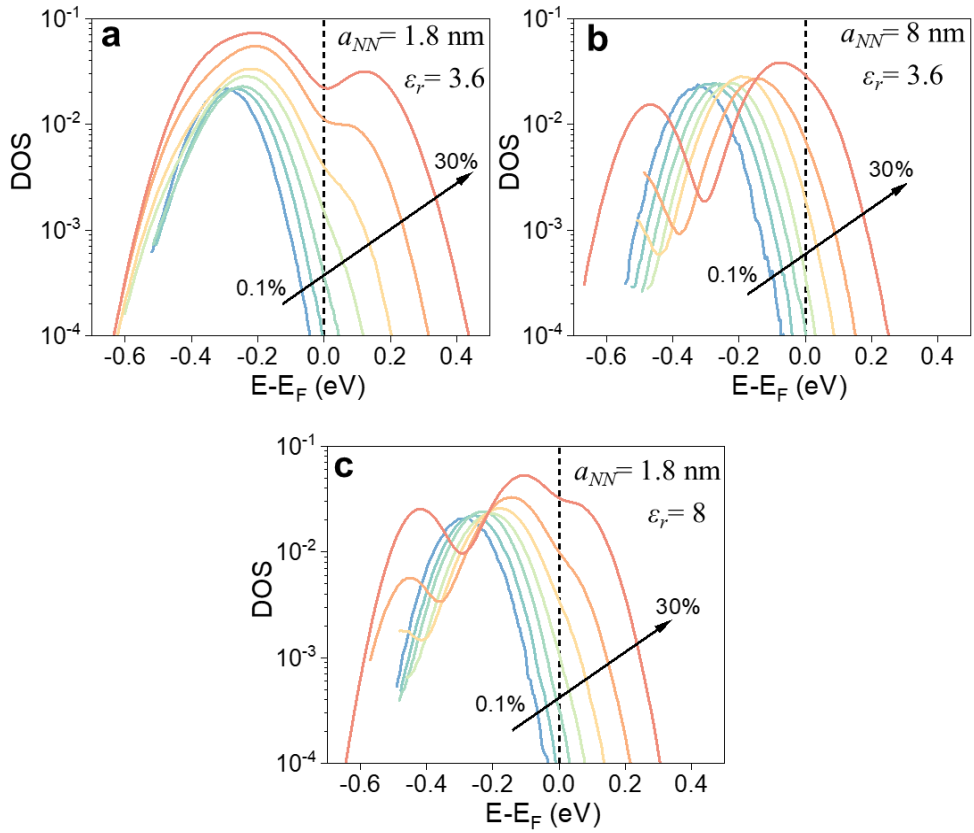

**Figure S23** | Evolution of DOS as a function of hole concentration from kMC simulation at different parameters. a) $a_{NN}=1.8$ nm, relative dielectric constant $\varepsilon_r=3.6$. b) $a_{NN}=8$ nm, $\varepsilon_r=3.6$. c) $a_{NN}=1.8$ nm, $\varepsilon_r=8$. The default kMC parameters: energetic disorder $\sigma_{DOS}=75$ meV, attempt-to-hop frequency $\nu_0=1\times10^{13}$ s$^{-1}$, temperature $T=300$ K. Hole concentration $c$ from 0.1% to 30%.

## S9-Thermoelectric characteristics of kMC simulation with different parameters.

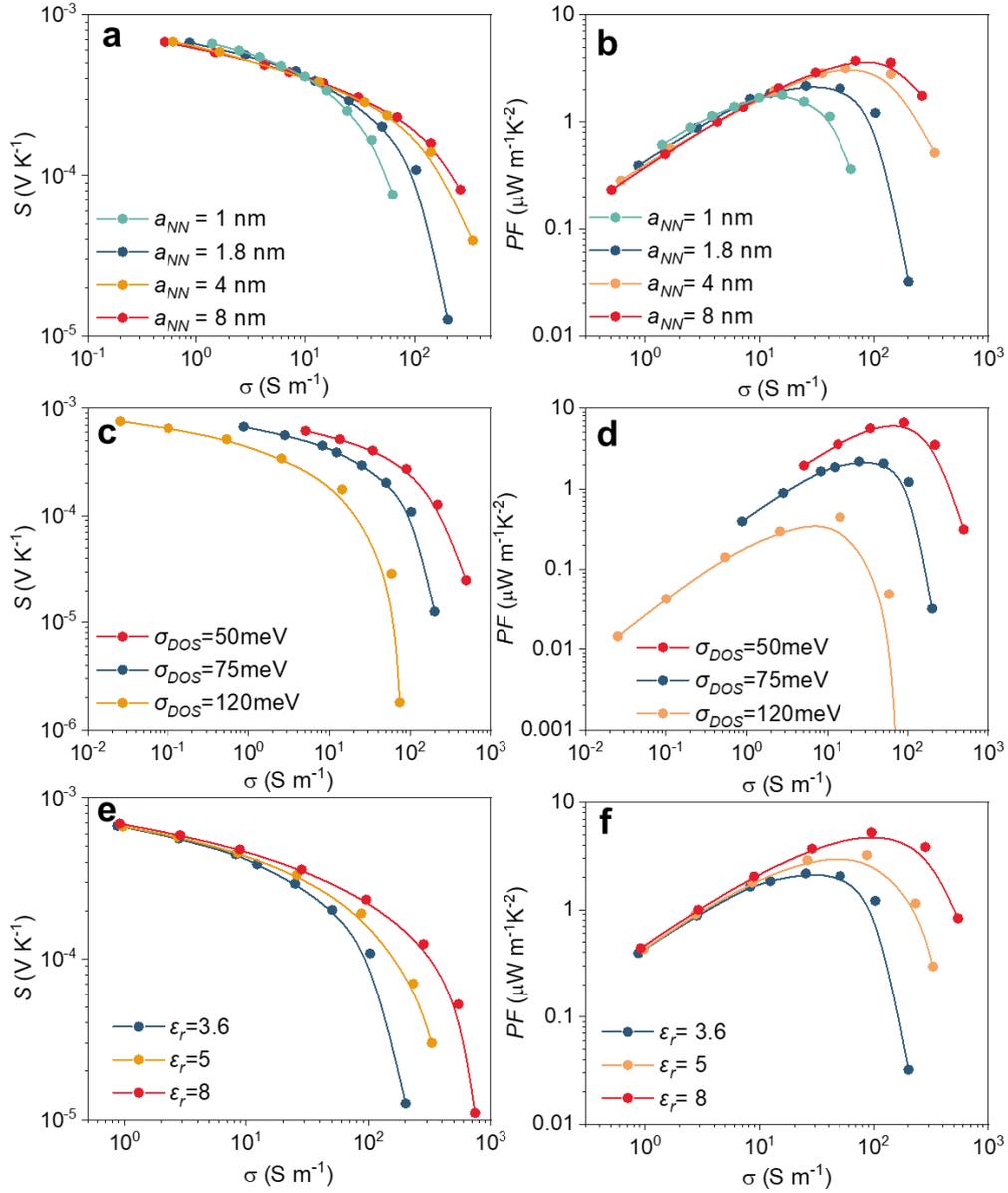

**Figure S24** | The change of thermoelectric characteristics by kMC simulation with different parameters. a, b) $a_{NN}$=1 nm, 1.8 nm, 4 nm, and 8 nm, respectively, energetic disorder $\sigma_{DOS}$= 75 meV, relative dielectric constant $\varepsilon_r$= 3.6. c, d) $a_{NN}$= 1.8 nm, $\sigma_{DOS}$= 50 meV, 75 meV, and 120 meV, respectively, $\varepsilon_r$= 3.6. e, f) $a_{NN}$= 1.8 nm, $\sigma_{DOS}$= 75 meV, $\varepsilon_r$= 3.6, 5, 8, respectively. Default kMC parameters: attempt-to-hop frequency $\nu_0$= 1×10$^{13}$ s$^{-1}$, temperature $T$ = 300 K. Hole concentration $c$ from 0.1% to 30%. These results show that a larger inter-site distance (localization length) and higher dielectric constant, which were shown above to effectively postpone the opening of a soft Coulomb gap, indeed facilitate improved charge transport and enhanced power factors by postponing the associated roll-off in PF. The enhancement of the power factor with reduced energetic disorder (panels c, d) is attributed to the increase in conductivity, as lower $\sigma_{DOS}$ facilitates more efficient charge transport.